\documentclass[onecolumn]{elsart}
\usepackage{epsf} 
\usepackage{rotating} 
\usepackage{rotate} 
\usepackage{natbib}
\usepackage[dvips]{psfrag}

\def\Rm{\mbox{Rm}} 
\def\u{\hbox{\bf u}} 
\def\U{\hbox{\bf U}}

\def\v{\hbox{\bf v}} 
 
\def\B{\hbox{\bf B}} 
\def\F{\hbox{\bf F}}

\def\E{\hbox{\bf E}} 
\def\A{\hbox{\bf A}}

\def\j{\hbox{\bf j}}

\theoremstyle{break}
\newtheorem {Prop}{Proposition}
\newtheorem {Rem}{Remark}

\def\vnabla{\mbox{\boldmath $\nabla$}} 
\def\zero{\mbox{\boldmath $0$}} 
  
\begin{document} 
 
\begin{frontmatter} 
 
\title {\mbox{Kinematic Dynamos using Constrained Transport}\\
with High Order Godunov Schemes\\and Adaptive Mesh Refinement} 
\author[saclay,iap]{Romain Teyssier}\thanks{E-mail addresses: Romain.Teyssier@cea.fr (R.Teyssier), s.fromang@qmul.ac.uk
(S.Fromang), dormy@phys.ens.fr (E.Dormy).}
\author[queenmary]{$\!\!$,\ S\'ebastien Fromang}
\author[lps,ipgpcnrs,iap]{$\!\!$,\ Emmanuel Dormy} 
\address[saclay]{CEA/DSM/DAPNIA/Service d'Astrophysique, Gif-sur-Yvette, 91191 Cedex, France.}
\address[iap]{Institut d'Astrophysique de Paris, 98$^{\mbox{\tiny ~bis}}$ Bd Arago, 75014 Paris, France.}
\address[queenmary]{Astronomy Unit, Queen Mary, University of London,
  Mile End Road,\\ London E1~4NS, U.K.}
\address[lps]{Laboratoire de Physique Statistique, E.N.S., 24, rue Lhomond\\
75231 Paris Cedex 05, France.}
\address[ipgpcnrs]{I.P.G. de Paris, France \& C.N.R.S., France.} 
 
\begin{abstract} 
We propose to extend the well-known MUSCL-Hancock scheme for Euler
equations to the induction  equation modeling the magnetic field  evolution
in kinematic dynamo problems.  The scheme is based on an integral form
of the underlying conservation law  which, in our formulation, results
in a ``finite-surface'' scheme for the induction equation.  This 
naturally leads to  the  well-known  ``constrained transport''  method, with
additional    continuity  requirement    on    the   magnetic    field
representation.   The  second ingredient in the    MUSCL scheme is the
predictor step  that ensures second order accuracy   both in space and
time.  We explore   specific constraints  that  the  mathematical
properties of the induction   equations place on this predictor  step,
showing that three possible variants can be considered. 
We  show  that the  most aggressive formulations (referred to as C-MUSCL
and U-MUSCL) reach  the same  level  of accuracy  as the
other one (referred to as Runge-Kutta), at a  lower computational cost.
More interestingly, these two
schemes  are  compatible  with  the  Adaptive Mesh   Refinement  (AMR)
framework. It has been implemented in the AMR code RAMSES. It offers a
novel  and efficient implementation of  a  second order scheme for the
induction equation.  We have tested it by solving two kinematic dynamo
problems in   the  low diffusion   limit.     The
construction  of this scheme for the  induction equation constitutes a
step towards solving the full MHD  set of equations using an extension
of our current methodology.
\end{abstract}  

\end{frontmatter}  

\section{Introduction}

The  extension  of  Godunov-type  conservative  schemes  for Euler
equations of fluid dynamics  \citep{Toro99,Bouchut05} to the system of
ideal magnetohydrodynamics (MHD)   has been   a matter of    intensive
research, starting from the early 90's. The great variety of different
MHD  implementations of the  original  Godunov method, especially in a
multidimensional setting, has left  several unexplored paths opened in
designing MHD conservative methods.

The most natural approach in adapting finite-volume schemes to the MHD
equations is to  define the magnetic field  component at the center of
each cell, where  the  traditional hydrodynamical variables  are  also
defined. One then takes advantage of decades of experience in the
development  of stable and  accurate shock-capturing schemes.  In this
case,   the solenoidality  constraint  $\vnabla\cdot\B=0$  has to be enforced
using either      a ``divergence cleaning''    step (see   for example
\citeauthor{Brackbill80},          \citeyear{Brackbill80}          and
\citeauthor{Ryu98},  \citeyear{Ryu98}),  or various  reformulations of
the   MHD   equations        including  additional    divergence-waves
\citep{Powell99} or divergence-damping terms \citep{Dedner02}. A novel
cell-centered  MHD    scheme   has    been   recently   developed   by
\cite{Crockett05} that  combines most of  these ideas  into one single
algorithm.

An   alternative approach is   to  use the Constrained Transport  (CT)
algorithm for the induction equation, as suggested in the late 60's by
\cite{Yee66},   and later revisited  by  \cite{EvansHawley88}. In this
description, the  magnetic field is defined  at the cell  faces, while
other hydrodynamical variables  are defined at  the cell center.  This
is often called a ``staggered mesh'' discretization.   As we will see
in this paper, CT provides a natural expression of the induction equation in
conservative form.  Combining CT  with the Godunov framework to design
high-order,  stable schemes is  therefore  a very attractive solution.
This  combined approach was  first explored in  the context of the MHD
equations   by   \cite{Balsara99}.    This   method   directly    uses
face-centered Godunov  fluxes and averages these on  the  cell edges to
estimate the  Electro-Motive Force   (EMF).  \cite{Toth00}  proposed  an
interesting    cell-centered  alternative to   this scheme.
More recently, \cite{Londrillo00, Londrillo04} have revisited 
the  problem   and shown   that  the  proper   way of  defining the
edge-centered EMF is to solve a 2D  Riemann problem at the cell edges.
They  have applied this  idea  to design high-order, Runge-Kutta,  ENO
schemes.  Finally, \cite{Gardiner05} have extended 
Balsara and Spicer scheme to
design a more stable and more robust way  of computing the EMF. 

The implementation of these  various schemes within the Adaptive  Mesh
Refinement framework is another challenging issue. It introduces two
main  new  technical   difficulties:  first, proper   fluxes  and  EMF
corrections  between different levels  of refinement must be accounted
for. Second, when   refining  or de-refining   cells,  divergence-free
preserving  interpolation   and     prolongation operators  must    be
designed. Both  of these  issues  have recently been discussed in the
framework   of the CT  algorithm by  several authors \citep{Balsara01,
Toth02, Li04}.

The purpose of this article is  to present a  novel algorithm based on a
high-order   Godunov  implementation of    the  CT algorithm within  a
tree-based     Adaptive Mesh  Refinement    (AMR)  code called  RAMSES
\citep{Teyssier02}.  As opposed to   the grid-based  (or  patch-based)
original     AMR   designed  introduced   by   \cite{Berger84}   and
\cite{Berger89}, tree-based  AMR  trigger local grid  refinements on a
cell by cell  basis. In this  way, the grid  follows  more closely the
geometrical features of  the computed flow, at the  cost of a greater
algorithm's complexity. Nevertheless, such tree-based AMR schemes have
been implemented  with success by  various authors in the framework of
astrophysics and      fluid dynamics  \citep{Kravtsov97,   Khokhlov98,
Teyssier02,  Popinet03} but not yet  in  the MHD  context. On  the other
hand, patch-based AMR    algorithms  have been developed  by   several
authors              in                 recent                   years
\citep{Balsara01,Kleimann04,Powell99,Samtaney04,Ziegler99}   and  used
for MHD applications. The main requirement that tree-based AMR usually
place on  the  underlying solver  is the  compactness of the computational
stencil: any high  order scheme with a stencil extending to two points, or
less,  in each direction  can  easily be coupled  to   an
``octree'' data structure 
\citep{Khokhlov98}.

In this paper, our goal is  to solve the  induction equation using the
MUSCL  scheme, originally presented  by  \cite{VanLeer77},  and widely
used in the literature for the Euler equations.  This very simple
method is second order  accurate in time and space  and has a  compact
stencil:  only  2 neighboring cells  in each  direction  (and for each
dimension) are necessary  to update the central  cell  solution to the
next time step.  This compactness property is of particular importance
for our  tree  based AMR approach. It   is also useful for an  efficient
parallelization  relying  on domain decomposition.   To our knowledge,
this is the first implementation of the MUSCL scheme combined with the
Constrained  Transport algorithm that  solves the  induction equation.
The key ingredient that ensures second order accuracy is the so-called
``predictor step'', in which the solution  is first advanced by half a
time step.  We will consider  a few different computational strategies
for this predictor step and discuss their respective merits.  Finally,
we will present our overall tree-based AMR scheme.

This paper  is limited to the  induction equation. We  intend to apply
the  same approach to  the full  MHD equations  in a future paper.
Nevertheless,  it is  interesting  to determine  if  such a  numerical
approach can  be applied to  kinematic dynamo problems, for  which the
induction equation  alone applies.  The induction  equation is linear,
but it can yield  remarkably rich magnetic instabilities corresponding
to   exponential   field   growth   and  referred   to   as   ``dynamo
instabilities''.   The  description  of  these instabilities,  and  the
conditions  under which  they  occur, constitute  an  active field  of
research,   with  important  consequences   in  astrophysics   and  in
geophysics, since  they account for  the origin of magnetic  fields in
the Earth,  planets, stars  and even galaxies.   We will  restrict our
attention here to well known dynamo flows and use them to investigate
the numerical properties of our scheme.

An important problem in dynamo theory  is related to a subclass of dynamo
flows, known as ``fast  dynamos'' which yield exponential field growth
with finite growth rates in  the limit of vanishing resistivity.  This
is  of  particular importance  for  astrophysical applications.   Fast
dynamos,  when  investigated   with  small,  but  finite,  resistivity
yield eigenmodes  that  are  very localized  in  space,  and are 
therefore ideal candidates for an investigation using the AMR scheme.

Dynamo problems have traditionally been studied using spectral methods
\citep{Galloway86,Benchmark01}.  Some recent models have been produced
using  finite   differences    \citep{Archontis03}, finite     volumes
\citep{Harder05} or  finite elements \citep{Matsui05}.  However, all of these
methods rely on  explicit  physical diffusion to  ensure numerical
stability.   The  interest of using  CT   within the Godunov framework
together with  an AMR approach is twofold.   First,  fast dynamo modes
have a very localized  spatial structure (scaling as $Rm^{-1/2}$ where
$Rm$  is the  magnetic  Reynolds number).   Adapting the computational
grid to the typical geometry  of  these modes  therefore appears as  a
very  natural strategy to  minimize  computational cost.   Second, the
Godunov  methodology, using   the  CT scheme,  introduces the  minimal
amount of numerical dissipation needed to ensure stability. This is an
important property when using an AMR approach, for which cells of very
different sizes   coexist.  This last  property of  the scheme is then
mandatory to allow the use of a coarse grid in regions barely affected
by the physical diffusion.

We will  present several tests that demonstrate the efficiency of our  
tree-based AMR Godunov CT scheme for solving complex dynamo problems:
we will first reproduce a simple advection problem of a magnetic loop
and then validate the approach on two well known dynamo flows: the
Ponomarenko dynamo and a fast ABC dynamo.

\section{Constrained Transport in Two Space Dimensions}
\label{section2}

In this section,  we  briefly review  the  design  of stable numerical
schemes for hyperbolic systems of conservation laws in two
space   dimensions    using   the   Godunov    approach.     Following
\cite{Londrillo00}, such systems are called here ``Euler systems'', as
opposed to the ``induction system'' we will consider later.

\subsection{First Order Godunov Scheme for Euler systems}
\label{euler}

We first  examine the  problem in one  space dimension.  The following
Euler system,
\begin{equation}
\frac{\partial \U}{\partial t} + \vnabla \cdot \F ( \U )  = 0  \, ,
\label{euler system}
\end{equation}
can  be written  in integral  form by  defining finite  control volume
elements    in   space and time,    where    we   define    a   cell    by
$V_i=[x_{i-\half},x_{i+\half}]$  and a  time interval  by $\Delta  t =
t^{n+1}-t^{n}$. The conservative system writes for each cell $V_i$
\begin{equation}
 \left< U \right>_i ^{n+1} - \left< U \right>_i ^{n}  + 
\frac{\Delta t}{\Delta x}
\left(  F _{i+ \half}^{n+\half} - F _{i-\half}^{n+\half} \right) = 0 \, .
\end{equation}
Note  that this  integral form  is exact  for the  corresponding Euler
system. The  averaged, cell-centered state is defined by
\begin{equation}
\left< U \right>_i ^n = \frac{1}{\Delta x} 
\int _{x_{i-\half}} ^{x_{i+\half}} U(x,t^n)\, {\rm d}x \, ,
\end{equation}
while the averaged, time-centered intercell flux is defined by
\begin{equation}
F_{i+\half}^{n+\half} = \frac{1}{\Delta t} 
\int _{t^n} ^{t_{n+1}} F(x_{i+\half},t)\, {\rm d}t \, .
\end{equation}

The Godunov  method states that  the intercell flux is  computed using
the solution of a Riemann problem  with left and right states given by
the left and right averaged states
\begin{equation}
U^*_{i+\half}(x/t) = RP\left[ \left< U \right> _i ^n,  
                              \left< U \right>_{i+1} ^n \right] \, .
\end{equation}

This  approach,   called  ``first  order   Godunov  scheme'',  assumes
that the solution  inside cell  $V_i$ is  {\it piecewise
constant}.   Taking advantage  of the  self-similarity of  the Riemann
solution for  initially piecewise  constant states, one  can simplify
further the time-average of the flux and obtain
\begin{equation}
F_{i+\half}^{n+\half} = F(U^*_{i+\half}(0)) \, .
\end{equation}

Note that again the time evolution  of the average state over one time
step is exact.  Numerical  approximations arise when one assumes {\it
at the next  time step} that the new  solution inside cell $V_i$
is also piecewise constant and equal to the new averaged state.

We now   extend the previous  method   to  Euler systems   in 2  space
dimensions.   The   conservative system can  also  be   written in the
following {\it unsplit} formulation
\begin{equation}
 \left< U \right>_{i,j} ^{n+1} - \left< U \right>_{i,j} ^{n}  + 
\frac{\Delta t}{\Delta x}
\left(  F _{i+ \half,j}^{n+\half} - F _{i-\half,j}^{n+\half} \right)  + 
\frac{\Delta t}{\Delta y}
\left(  G _{i,j+ \half}^{n+\half} - G _{i,j-\half}^{n+\half} \right)  =0 \, ,
\end{equation}
where  the average  state is  now defined  over a  2  dimensional cell
$V_{i,j}$,  and  intercell  fluxes  are  now  time  averaged  fluxes
integrated over the line separating neighboring cells
\begin{equation}
F_{i+\half,j}^{n+\half} = 
\frac{1}{\Delta t} 
\frac{1}{\Delta y} 
\int _{t^n} ^{t_{n+1}} 
\int _{y_{j-\half}} ^{y_{j+\half}} 
F(x_{i+\half},y,t)\, {\rm d}t\, {\rm d}y \, ,
\end{equation}
\begin{equation}
G_{i,j+\half}^{n+\half} = 
\frac{1}{\Delta t} 
\frac{1}{\Delta x} 
\int _{t^n} ^{t_{n+1}} 
\int _{x_{i-\half}} ^{x_{i+\half}} 
G(x,y_{j+\half},t)\, {\rm d}t\, {\rm d}x \, .
\end{equation}
At  this point,  the integral  form is  still  exact.  The
generalization of  the 1D Godunov scheme  to multidimensional problems
now relies on solving two dimensional Riemann problems at each corner,
defined by four initially piecewise constant states
\begin{equation}
U^*_{i+\half,j+\half}(x/t,y/t) = RP
\left[ 
\left< U \right> _{i,j} ^n,  
\left< U \right>_{i+1,j} ^n 
\left< U \right>_{i,j+1} ^n 
\left< U \right>_{i+1,j+1} ^n 
\right] \, .
\end{equation}

The fundamental  difference with the  1D case is  that we now  need to
average the  complete solutions of  2 adjacent Riemann  solutions over
the entire  transverse line segment, where fluxes  are defined.  These
space-averaged  fluxes  are  not  functions of  a  unique  self-similar
variable  anymore, but depend  explicitly on  time.  Building  such a
numerical scheme is barely possible for simple scalar linear advection
problem and far  too complex  to implement  for  non-linear systems.

The traditional approach  is to approximate the true  solution using a
predictor-corrector  scheme. This is  also the  key ingredient  of any
high-order scheme,  where the  self-similarity of the  Riemann problem
breaks  down, even  in  one  space dimension,  due  to the  underlying
piecewise linear or parabolic representation of the data. The idea is
to compute a predicted state at time level $t^{n+1/2}$ and to use this
intermediate state as an input state for the two final 1D Riemann solvers.

We list here 3 classical methods to implement this predictor step
\begin{itemize}
\item  {\bf Godunov  method}: no  predictor step  is  performed.  This
  greatly  simplifies the  method,  which now  relies  on one  Riemann
  solver in each  direction. The prize to pay  is a somewhat restrictive
  Courant  stability condition: $(u/\Delta  x+v/\Delta y)\Delta  t \le
  1$, where $u$ and $v$ are the maximum wave speed in each direction.
\item {\bf Runge-Kutta method}:  the predictor step is performed using
  the  2D Godunov  method  with  half the  time  step.  The  resulting
  intermediate  states are  then used  to compute  the fluxes  for the
  final conservative update. The Courant  condition is the same as for
  the Godunov method, but one  has to perform 2 Riemann solvers per
  cell in each direction (4 in total).
\item  {\bf Corner Transport  Upwind method}:  predicted states  for a
  given  Riemann  problem  are  computed  with  a  1D  update  in  the
  transverse direction only, for the time interval $\Delta t/2$.  This
  scheme was first proposed by \cite{Colella90}. It allows up to a factor
  of two larger
  time  steps  than  the  two  previous  schemes,  since  the  Courant
  condition is now  $\max (u/\Delta x,v/\Delta y)\Delta t  \le 1$, but
  2 Riemann solvers per cell in each direction (4 in total) are still
  needed.
\end{itemize}
All   three   methods   are   directionally   unsplit,   first   order
approximations (in space) of the underlying Euler system.

\subsection{First Order Godunov Scheme for the Induction Equation}

The magnetic field  evolution in the MHD approximation  is governed by
the  induction  equation  which   neglects  free  charge  density  and
displacement currents. It is written in conservative form as
\begin{equation} 
\frac{\partial \B}{\partial t} = \vnabla \times \E  +  
\eta \,\Delta \B \, ,  
\label{magnetic induction eq} 
\end{equation} 
where the EMF $\E$ is given by
\begin{equation} 
\E = \v \times \B \, ,
\label{emf} 
\end{equation} 
and $\eta$ is the  magnetic diffusivity.  The magnetic field also
satisfies the divergence free constraint
\begin{equation} 
\vnabla \cdot \B = 0  \, .
\label{div-free cond} 
\end{equation} 
It is usually more convenient to consider (\ref{magnetic induction eq})
in non-dimensional form by introducing a typical lengthscale ${\mathcal L}$ 
and a typical timescale ${\mathcal T}={\mathcal L}/U$ where $U$ is some norm 
of the velocity (usually based on the maximal value over space and time).
The resulting non-dimensional equation is
\begin{equation} 
\frac{\partial \B}{\partial \tilde{t}} = \vnabla \times \left( \tilde{\v} \times \B  \right)  
+  \Rm^{-1} \,\, \Delta \B \, ,  
\label{magnetic induction eq adim} 
\end{equation} 
where $\Rm = (U {\mathcal L})/\eta$ while $\tilde{t}=t/{\mathcal T}$ 
and $\tilde{\v}=\v/U$ are respectively the non-dimensionnal time and 
velocities and the spatial derivative are taken with respect to 
normalized distances.

The  EMF $\E$  is  here the  analog  of the  flux  function for  Euler
systems.     We   now   restrict    our attention  to    2D   dimensional
flows\footnote{The   one    dimensional   induction   equation,   with
$B_x=$constant, is  equivalent to a  Euler system,  for which
the  standard methodology  applies without  modification.},  for which
only one component of the EMF, say $E_z$, is sufficient.

Following the Godunov approach, we  write the  2D induction  equation in
integral  form  over a  finite  control  volume  in space and time.   For
the $B_x$ component  of the  magnetic field, we  define a  finite surface
element    $S_{i+1/2,j}=[y_{j-1/2},y_{j+1/2}]$    at    position
$x_{i+1/2}$ 
\begin{equation}
 \left< B_x \right>_{i+\half,j} ^{n+1} = \left< B_x \right>_{i+\half,j} ^{n} +
\frac{\Delta t}{\Delta y}
\left(  \left< E_z \right>_{i+ \half,j+\half}^{n+\half} - 
\left< E_z \right>_{i+\half,j-\half}^{n+\half} \right) \, .
\label{induction_num_x}
\end{equation}

For  the $B_y$ component,   we   define  a   finite   surface   element
$S_{i,j+1/2}=[x_{i-1/2},x_{j+1/2}]$  at  position  $y_{i+1/2}$. The  
induction equation  in integral form has a similar representation
\begin{equation}
 \left< B_y \right>_{i,j+\half} ^{n+1} = \left< B_y \right>_{i,j+\half} ^{n} -
\frac{\Delta t}{\Delta x}
\left(  \left< E_z \right>_{i+ \half,j+\half}^{n+\half} - 
\left< E_z \right>_{i-\half,j+\half}^{n+\half} \right) \, .
\label{induction_num_y}
\end{equation}

Note  that this  integral form  in space and time  is exact.  The average,
surface centered, magnetic states  are defined as the average magnetic
field components on their corresponding control surfaces
\begin{equation}
  \left< B_x \right>_{i+\half,j}^n = \frac{1}{\Delta y}
  \int _{y_{i-\half}} ^{y_{i+\half}} B_x(x_{i+\half},y,t^n)\, {\rm d}y \, ,
  \label{bxdef}
\end{equation}
\begin{equation}
  \left< B_y \right>_{i,j+\half} ^n= \frac{1}{\Delta x}
  \int _{x_{i-\half}} ^{x_{i+\half}} B_y(x,y_{i+\half},t^n)\, {\rm d}x \, .
 \label{bydef}
\end{equation}

\subsubsection{2D Riemann Problem}

The time  centered EMF results from a time  average at  the corner
points
\begin{equation}
\left< E_z \right>_{i+\half,j+\half}^{n+\half} = 
\frac{1}{\Delta t} 
\int _{t^n} ^{t_{n+1}} E_z(x_{i+\half},y_{j+\half},t)\, {\rm d}t \, .
\label{EMF function}
\end{equation}

Let us now apply the  Godunov method  to the  2D induction  equation. Upon 
noticing  that  our initial  conditions are given by four piecewise
constant states around each corner points, we can use the self-similar
solution of the 2D Riemann problem  at the corner point,
\begin{equation}
U^*_{i+\half,j+\half}(x/t,y/t) = RP
\left[ 
\left< U \right> _{i,j} ^n,  
\left< U \right>_{i+1,j} ^n 
\left< U \right>_{i,j+1} ^n 
\left< U \right>_{i+1,j+1} ^n 
\right] \, ,
\end{equation}
and  time integration vanishes in equation~(\ref{EMF function})
\begin{equation}
\left< E_z \right>_{i+\half,j+\half}^{n+\half} = 
E_z(U^*_{i+\half,j+\half}(0,0)) \, .
\end{equation}

The Godunov method, applied to the induction equation in 2D, shares this
interesting  property with  the  Godunov method  applied  to 1D  Euler
system. The self-similarity of the flux function was lost for 2D Euler
systems. The  self-similarity of the  EMF function is still  valid for
the 2D induction equation, provided  our initial conditions are described by
piecewise constant states. We will  see in the next section, that this
is unfortunately not true in the general case, even at lowest order.

\begin{figure}
\psfrag{ijk}[][]{\small $(i,j,k)$}
\psfrag{ijpk}[][]{\small $(i,j\!+\!1,k)$}
\psfrag{ipjpk}[][]{\small \ \ \ $(i\!+\!1,j\!+\!1,k)$}
\psfrag{ipjk}[][]{\small \ $(i\!+\!1,j,k)$}
\psfrag{x}[][]{\footnotesize {\bf x}}
\psfrag{y}[][]{\footnotesize {\bf y}}
\centerline{
\includegraphics[width=6cm]{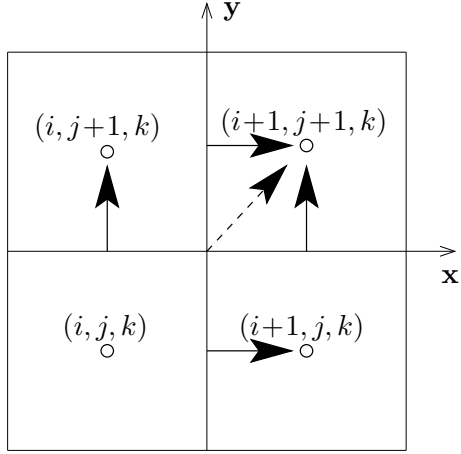}
}
\caption{The 2D Riemann problem in the x-y plane to compute the EMF 
in the z direction at edge ($i+\half$, $j+\half$). The face-centered 
magnetic fields are shown as vertical and horizontal arrows.  The
velocity field is shown as the dashed arrow.}
\label{2DRP}
\end{figure}

As noticed  by \cite{Londrillo00}, the  2D Riemann problem is  the key
ingredient for  solving the induction equation with  a stable (upwind)
scheme. The  4 initial  states (with 2  magnetic field  components per
state)  need to satisfy  the $\vnabla  \cdot \B  = 0$  property.  
$B_x$ should therefore be the same  for the two top  states, and for
the two bottom states, while $B_y$ should be the same for the two left
states,  and for  the  two right  states  (see Fig.~\ref{2DRP}).  This
condition is naturally satisfied as  long as magnetic field is defined
as a surface-average, see (\ref{bxdef}) and (\ref{bydef}).

In  the   general  MHD  case,  designing  2D   Riemann  solvers (even
approximate  ones)  is  a  very  ambitious  task.  For  the  kinematic
induction  case,  the  solution  is  however remarkably  simple,  since  the
solution is nothing  else but the upwind state.  The edge-centered EMF
can therefore be written in the following closed form
\begin{eqnarray}
\nonumber
\left< E_z \right>_{i+\half,j+\half}^{n+\half} = 
         u \frac{ \left< B_y \right>_{i+1,j+\half}+\left< B_y \right>_{i,j+\half} }{2} 
     -   v \frac{ \left< B_x \right>_{i+\half,j+1}+\left< B_x \right>_{i+\half,j} }{2} \\
-\left| u \right|  \frac{ \left< B_y \right>_{i+1,j+\half}-\left< B_y \right>_{i,j+\half} }{2} 
+\left| v \right| \frac{ \left< B_x \right>_{i+\half,j+1}-\left< B_x \right>_{i+\half,j} }{2} 
 \, ,
\label{2DRPsol}
\end{eqnarray}
where $u$ and  $v$ are respectively the $x$ and $y$  components of the flow
velocity   $\v   =  (u,v,w)$   computed   at   the   center  of the edge
$(i+\half,j+\half)$.  This  last   equation  is  familiar  in  the
framework of upwind finite-volume  schemes. It can be decomposed
into two contributions.  The first line is the  EMF computed using the
average  magnetic  fields   at  the  cell  corners:  this   EMF  is  a
second-order  in  space. The  resulting  scheme  (retaining this term only)
would have  been 
unconditionally  unstable, if  it was  not  for the  second term,  the
contribution of  the upwinding.   It is equivalent  to a  2D numerical
diffusivity,  with  directional   diffusivity  coefficients  given  by
$\eta_x = \left| u\right| \Delta x / 2$ and $\eta_y = \left| v \right|
\Delta y  / 2$.  This (relatively large) resistivity introduces  the minimal but
necessary amount of numerical diffusion for the scheme to remain stable.

\subsubsection{Constrained Transport as a Finite Surface Approximation} 

This straightforward extension of  the Godunov methodology has lead us
to the well known  ``Constrained Transport'' (CT) scheme, that
was designed a  long time ago for the  MHD equations by \cite{Yee66}. The
key property of the CT scheme  is that one can also write the $\vnabla
\cdot \B =0$ constraint in integral form as
\begin{equation}
\frac{ \left< B_x \right>_{i+\half,j} ^n - \left< B_x \right>_{i-\half,j} ^n}
{\Delta x} +
\frac{ \left< B_y \right>_{i,j+\half} ^n - \left< B_y \right>_{i,j-\half} ^n }
{\Delta y} = 0 \, .
\label{divBint}
\end{equation}

This  integral form is  exact.  Moreover,  if it  is satisfied  by our
initial  data, the  integral forms  in (\ref{induction_num_x})
and  (\ref{induction_num_y}) ensure  that it  will be  satisfied  at all
iterations during the numerical  integration.  Using 
equation~(\ref{divBint}), and assuming that formally 
$\Delta x \rightarrow 0$, we show that the following property holds:
\begin{Rem}
$\left< B_x \right>_{j} ^n (x) $ is a continuous function of coordinate $x$,
\end{Rem}
and, symmetrically, assuming that formally $\Delta y \rightarrow 0$, we
have:
\begin{Rem}
$\left< B_y \right>_{i} ^n (y)$ is a continuous function of coordinate $y$,
\end{Rem}

This means that $\left< B_x  \right>_{i+1/2,j}^n$ can be considered as
piecewise constant in  the y direction, but {\it  has to be considered
as  piecewise linear in the x  direction}. This constitutes our lowest
order approximation  of the magnetic  field.  Symmetrically, to lowest
order, $\left< B_y \right>_{i,j+1/2}^n$ can be considered as piecewise
constant   in  the x direction,  but   {\it has   to  be considered as
piecewise linear in the y direction}.\footnote{Let us stress that for 
ideal MHD, a jump perpendicular to the fieldline is allowed.}

This last property provides a fundamental difference between the induction
equations  and  Euler systems.   It  is  due  to the  divergence  free
constraint, expressed  in integral form on a  staggered magnetic field
representation. One  consequence of this property is  that our initial
state for the 2D Riemann  problem cannot be piecewise constant anymore,
but  instead piecewise  linear. We therefore  loose  the property  of
self-similarity for the Riemann  solution at corner points, and cannot
perform an exact time integration to compute the time average EMF.  We
now have   to   rely   on  approximations.  
Following the strategies developed in section~\ref{euler}, we
approximate  the time  averaged EMF  using  various predictor-corrector
schemes.

\subsection{The Predictor step}
\subsubsection{Godunov Scheme}

The  first possibility is  to drop  the predictor  step and  solve the
Riemann     problem      defined     at     time      $t^n$.     Using
(\ref{induction_num_x})  and  (\ref{induction_num_y}),  together
with  the  EMF computed  from  (\ref{2DRPsol}),  we obtain  the
Godunov scheme  for the  induction equation. In  the simple case  of a
constant  velocity field  with  $u>0$ and  $v>0$  (the pure  advection
case), we can write the overall scheme as 
\begin{eqnarray}
 \nonumber
 \left< B_x \right>_{i+\half,j} ^{n+1} = \left< B_x \right>_{i+\half,j} ^{n} 
&+& u \frac{\Delta t}{\Delta y}
\left(  \left< B_y \right>_{i,j+\half}^n - \left< B_y \right>_{i,j-\half}^n \right) \\
&-& v \frac{\Delta t}{\Delta y}
\left(  \left< B_x \right>_{i+\half,j}^n - \left< B_x \right>_{i+\half,j-1}^n \right) 
\, .
\label{godunov_induction}
\end{eqnarray}
Using the $\vnabla  \cdot \B =0$ constraint at  time $t^n$ in integral
form (\ref{divBint}), we further simplify the scheme to obtain
\begin{eqnarray}
 \nonumber
 \left< B_x \right>_{i+\half,j} ^{n+1} = \left< B_x \right>_{i+\half,j} ^{n} 
&-& u \frac{\Delta t}{\Delta x}
\left(  \left< B_x \right>_{i+\half,j}^n - \left< B_x \right>_{i-\half,j}^n \right) \\
&-& v \frac{\Delta t}{\Delta y}
\left(  \left< B_x \right>_{i+\half,j}^n - \left< B_x \right>_{i+\half,j-1}^n \right) 
\, .
\label{godunov_advection}
\end{eqnarray}
We can therefore conclude:
\begin{Prop}
For the advection case, if  the initial data satisfy the integral form
of the solenoidality constraint,  the Godunov method for the induction
equation is identical to the Godunov method for the advection equation
on the staggered grid.
\end{Prop}

This rather simple point is  actually quite important, since it proves
that CT has advection properties quite similar
(in  this case  identical) to  traditional finite-volume  methods. The
Godunov scheme  for the induction  equation has a compact  stencil. It
is however of mere theoretical
interest, since,  as we will  see in the  next section, it is  not the
first  order  limit of  higher  order  Godunov  implementations of  the
induction equation.

\begin{figure}
\centerline{$\!\!\!$Runge-Kutta$\!\!\!$\hskip2.9cm U-MUSCL\hskip2.9cm C-MUSCL}
\centerline{
\epsfxsize 4.3cm
\epsffile{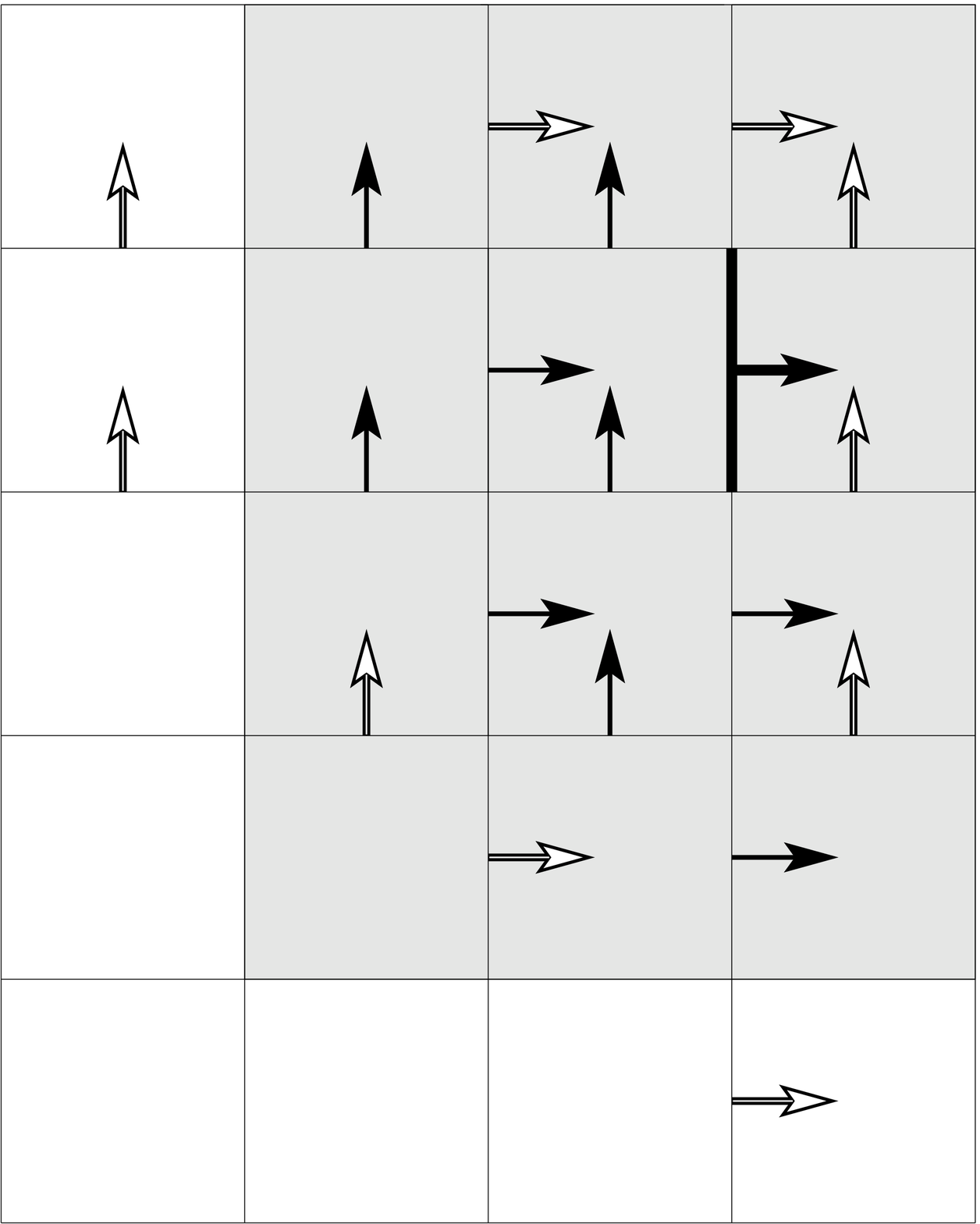}
\hskip 3mm
\epsfxsize 4.3cm
\epsffile{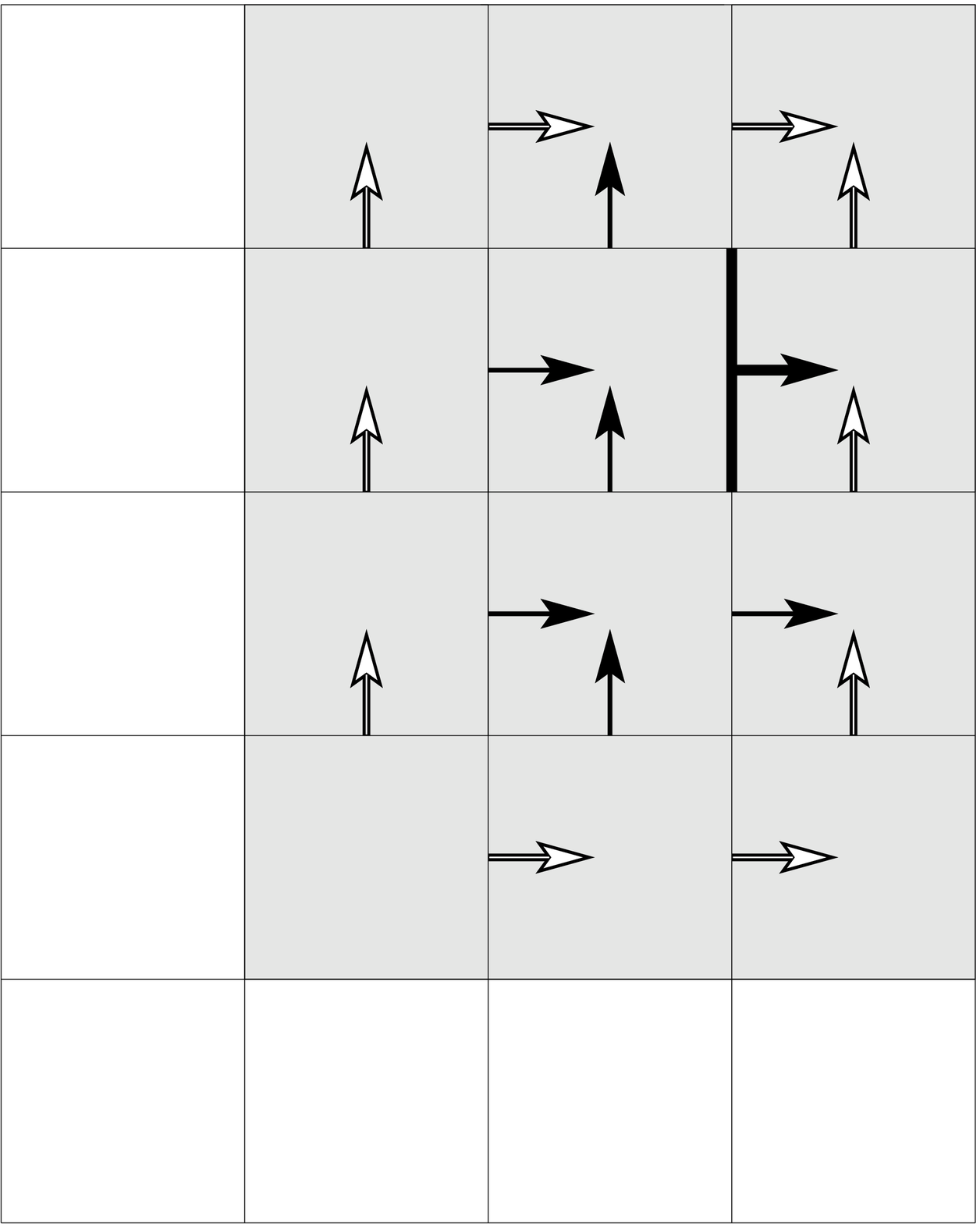}
\hskip 3mm
\epsfxsize 4.3cm
\epsffile{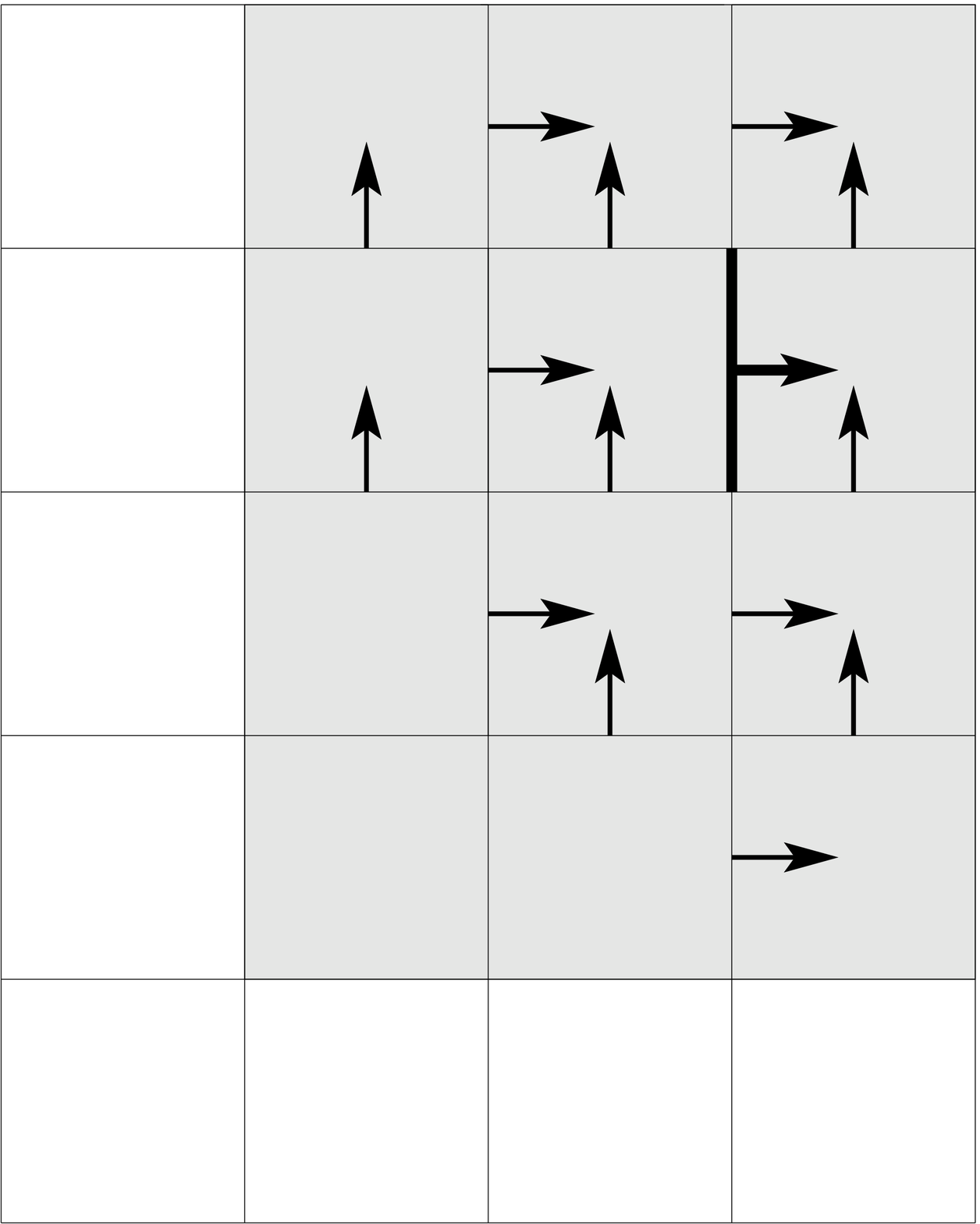}
}
\caption{Stencils of our various schemes for the induction
equation: Runge-Kutta scheme (left plot), U-MUSCL scheme (middle plot)
and C-MUSCL scheme (right plot).  The flux being computed is indicated
by  a bold  face and arrow.  For the  purpose of this  example, the
velocity field  is pointing  in the upper  right direction  ($u>0$ and
$v>0$).   The first order  stencil in space (second order in time) is  
represented with  black arrows.
Additional components required for the second order stencils in time and space 
are shown
with  white  arrows.   The  shaded  region indicates  cells  that  are
available  in a  tree-based AMR  implementation.  Only  the  two right
schemes have stencils compact enough for such an implementation.  }
\label{figstencils}
\end{figure}

\subsubsection{Runge-Kutta Scheme}

As discussed above, the $\vnabla \cdot \B =0$ constraint, and the loss
of  self-similarity in the  Riemann solution,  pushes towards  using a
predictor step in  designing our first order scheme.  The most natural
approach is the Runge-Kutta scheme, for which the solution is advanced
first  to  the intermediate  time  coordinate  $t^{n+1/2}$, using  the
(previously   described)  Godunov  scheme   with  time   step  $\Delta
t/2$. These  predicted states  are then used  to define the  4 initial
states for the  2D Riemann problem. The resulting EMF  is used to
advance  the solution  from time  $t^n$ to  the next time coordinate
$t^{n+1}$ with  time step $\Delta  t$. A similar, 2  step, Runge-Kutta
method   for  the   induction  equation   is  used   for   example  in
\cite{Londrillo00} and \cite{Londrillo04} to solve the MHD equations.

Using similar  arguments as in the  previous section, it  is easy to
show that, for a uniform  velocity field, since the predicted magnetic
field satisfies the integral form of the solenoidality constraint, the
corrector  step  for  the  induction  equation  is  identical  to  the
predictor step for the advection  equation. As we have shown in
the last section, this property  also holds for the predictor step, we
therefore obtain a second important result:

\begin{Prop}
For a uniform velocity field, if the initial data satisfy the integral
form of  the solenoidality constraint, the Runge-Kutta  method for the
induction  equation is  identical to  the Runge-Kutta  method  for the
advection equation on the staggered grid.
\end{Prop}

We will  show later that  it is also  possible to design  higher order
schemes for this  algorithm.  This scheme has two  nice properties: it
is second  order in time (while  still first order in  space), and the
predicted magnetic  field satisfies exactly  $\vnabla \cdot \B^{n+1/2}
=0$. There are also issues associated with it, especially in  the AMR
framework. It
can be  easily shown (see Fig.~\ref{figstencils}) that  the stencil is
not compact enough  for a tree-based AMR: 3 ghost  cells are needed in
each direction  (resp.  2) for  the second order (resp.   first order)
scheme. We will see in the  test section that it is also slightly more
diffusive than  the other  schemes we will describe in  the following
sections. The Courant stability condition is also rather restrictive
\begin{equation}
\left( \frac{u}{\Delta x} + \frac{v}{\Delta y} \right) \Delta t \le 1 \, .
\end{equation}

\subsubsection{Upwind-MUSCL Scheme}

When  deriving the  MUSCL scheme  for Euler  systems, \cite{VanLeer77}
noticed  that  it was  not  necessary for  the  predictor  step to  be
strictly  conservative.  A conservative  update was  however mandatory
for the corrector step.  Similarly,  for the induction equation, it is
{\it a  priori} not  necessary for the  predictor step to  satisfy the
solenoidality constraint. It is  however mandatory for the initial and
final data. Instead of computing one  EMF at each cell corner, using a
2D Riemann solver,  we now propose  to compute  {\it for the predictor
step  only} 4 EMFs  at each  cell corner,  corresponding to  each input
magnetic field.

These  EMFs  are  defined  as $\left<  E_z  \right>_{i+1/2,j+1/2}^{L}$,
$\left<      E_z      \right>_{i+1/2,j+1/2}^{R}$,     $\left<      E_z
\right>_{i+1/2,j+1/2}^{B}$ and $\left< E_z \right>_{i+1/2,j+1/2}^{T}$,
where each  upper index corresponds  to the ``left'',  ``right'', ``bottom''
and  ``top''  face,  respectively.  Each  EMF is  specialized  to  its
corresponding face-centered magnetic field component. One EMF per face
is allowed, in order to  satisfy the continuity constraint: we need to
solve  a  1D Riemann  problem  in  the  perpendicular direction.   The
Riemann solution  is here the  ``upwind'' state. The ``bottom''  and ``top''
EMF for the predictor step are therefore
\begin{eqnarray}
\nonumber
\left< E_z \right>_{i+\half,j+\half}^{B} &=& 
 u \left({ \left< B_y \right>_{i+1,j+\half}+\left< B_y \right>_{i,j+\half} }\right)/{2}
-v \left< B_x \right>_{i+\half,j} \\
\nonumber
&-&\left|  u  \right|  \left({ \left< B_y \right>_{i+1,j+\half}-
\left< B_y \right>_{i,j+\half} }\right)/{2} \, ,\\
\nonumber
\left< E_z \right>_{i+\half,j+\half}^{T} &=&
 u \left({ \left< B_y \right>_{i+1,j+\half}+\left< B_y \right>_{i,j+\half} }\right)/{2}
-v \left< B_x \right>_{i+\half,j+1} \\
&-&\left|  u  \right|  \left({ \left< B_y \right>_{i+1,j+\half}-
\left< B_y \right>_{i,j+\half} }\right)/{2} \, .
\label{1DRPsol_bottom_top}
\end{eqnarray}
Similarly, the ``left'' and ``right'' EMF are
\begin{eqnarray}
\nonumber
\left< E_z \right>_{i+\half,j+\half}^{L} =
 u \left< B_y \right>_{i,j+\half} 
&-&v \left({ \left< B_x \right>_{i+\half,j+1}+\left< B_x \right>_{i+\half,j} }\right)/{2} \\
\nonumber
&+&\left| v \right| \left({ \left< B_x \right>_{i+\half,j+1}-
\left< B_x \right>_{i+\half,j} }\right)/{2} \, ,\\
\nonumber
\left< E_z \right>_{i+\half,j+\half}^{R} = 
 u \left< B_y \right>_{i+1,j+\half} 
&-&v \left({ \left< B_x \right>_{i+\half,j+1}+\left< B_x \right>_{i+\half,j} }\right)/{2} \\
&+&\left| v \right| \left({ \left< B_x \right>_{i+\half,j+1}-
\left< B_x \right>_{i+\half,j} }\right)/{2} \, .
\label{1DRPsol_left_right}
\end{eqnarray}
The predictor step for the x component of the magnetic field becomes
\begin{equation}
 \left< B_x \right>_{i+\half,j} ^{n+1/2} = \left< B_x \right>_{i+\half,j} ^{n} +
\frac{\Delta t}{2\Delta y}
\left(  \left< E_z \right>_{i+ \half,j+\half}^{B} - 
\left< E_z \right>_{i+\half,j-\half}^{T} \right) \, ,
\label{predictor_x}
\end{equation}
and for the y component we have
\begin{equation}
 \left< B_y \right>_{i,j+\half} ^{n+1/2} = \left< B_y \right>_{i,j+\half} ^{n} -
\frac{\Delta t}{2\Delta x}
\left(  \left< E_z \right>_{i+ \half,j+\half}^{L} - 
\left< E_z \right>_{i-\half,j+\half}^{R} \right) \, .
\label{predictor_y}
\end{equation}
To complete this scheme, the corrector step is performed using
a  final   2D  Riemann  solver   to  compute  the   time-centered  EMF
(\ref{2DRPsol}) and  a final conservative update  of each magnetic
field          component         (\ref{induction_num_x})         and
(\ref{induction_num_y}).

Let us now examine the property of the Upwind--MUSCL scheme in the
case of a uniform
velocity field.  We can assume, without loss of generality, that $u>0$ and
$v>0$. In  this case,  the predicted  state can be  written in  a more
compact form
\begin{equation}
 \left< B_x \right>_{i+\half,j} ^{n+1/2} = \left< B_x \right>_{i+\half,j} ^{n} + 
 u \frac{\Delta t}{2\Delta y}
 \left(  \left< B_y \right>_{i,j+\half}^{n} - 
\left< B_y \right>_{i,j-\half}^{n} \right) \, ,
\end{equation}
which is equivalent, using (\ref{divBint}), to  
\begin{equation}
 \left< B_x \right>_{i+\half,j} ^{n+1/2} = \left< B_x \right>_{i+\half,j} ^{n} - 
 u \frac{\Delta t}{2\Delta x}
 \left(  \left< B_x \right>_{i+\half,j}^{n} - 
\left< B_x \right>_{i-\half,j}^{n} \right) \, .
\end{equation}
Similar expressions can be  derived for $ \left< B_y \right>_{i,j+1/2}
^{n+1/2}$.      Inserting      these     predicted     values     into
(\ref{2DRPsol}) and (\ref{induction_num_x}),  we get, after some
tedious manipulations, the final updated solution
\begin{eqnarray}
\nonumber
\left< B_x \right>_{i+\half,j} ^{n+1}&=&
\left< B_x \right>_{i+\half,j} ^{n}\left( 1-C_x\right) \left(1-C_y \right)+
\left< B_x \right>_{i-\half,j} ^{n}C_x\left(1-C_y \right)\\
&+&\left< B_x \right>_{i+\half,j-1} ^{n}C_y\left( 1-C_x\right)+
\left< B_x \right>_{i-\half,j-1} ^{n}C_xC_y \, ,
\label{ctu2d}
\end{eqnarray}
where  the following definitions have been used $C_x=u\Delta t/\Delta  x$  and  $C_y=v\Delta
t/\Delta y$. One can recognize here the Corner Transport Upwind (CTU) advection scheme
presented  in  \cite{Colella90},   for  which  the  Courant  stability
condition is
\begin{equation}
\max \left[ \frac{u}{\Delta x},\frac{v}{\Delta y} \right] \Delta t \le 1 \, .
\label{cfl}
\end{equation}
We therefore conclude:
\begin{Prop}
For a uniform velocity field, if the initial data satisfy the integral
form of  the solenoidality  constraint, the {\it  Upwind-MUSCL Scheme}
for the induction  equation is identical to Colella's  first order CTU
scheme for the advection equation on the staggered grid.
\end{Prop}

It is apparent in (\ref{ctu2d}) that the  stencil of this MUSCL scheme
is more   compact that  it is  for  the Runge-Kutta  scheme (see  also
Fig.~\ref{figstencils}). Since our goal is here to develop an AMR code
for the  induction equation, this is  a very attractive solution.  The
predictor step  is performed using  upwinding in the normal direction.
As for  Colella's CTU scheme, the  Courant stability condition is very
efficient.    We  now  explore one  last   possibility for   our MUSCL
predictor step.
 
\subsubsection{Conservative-MUSCL Scheme}
\label{sectCMUSCL}
The last scheme was  designed in dropping the solenoidality constraint
for  the  predictor step.  We  propose in  this  section  to drop  the
upwinding in  the EMF  computation for the  predictor step,  which now
becomes
\begin{equation}
\nonumber
\left< E_z \right>_{i+\half,j+\half}^{n} = 
 u \frac{ \left< B_y \right>_{i+1,j+\half}^{n} +\left< B_y \right>_{i,j+\half}^{n}  }{2} 
-v \frac{ \left< B_x \right>_{i+\half,j+1}^{n} +\left< B_x \right>_{i+\half,j}^{n}  }{2} \, .
\label{CMUSCL}
\end{equation}
Since we now have a single EMF per cell corner, the predicted magnetic
field satisfies  by construction  $\vnabla \cdot \B^{n+1/2}  =0$.  The
corrector step is the same as  for all 3 methods. Here again, we would
like to  examine the property  of the scheme  for the case  of uniform
advection.   Because  in this  case $\vnabla  \cdot \B^{n+1/2}
=0$, the  corrector step  is identical to  the corrector step  for the
Godunov advection  scheme on the staggered grid.   The predictor step,
on the other hand, can be  written as the Forward Euler scheme for the
advection equation  on the staggered grid. When  combined together, we
obtain  a new  first  order  advection scheme  for  which the  Courant
stability condition is the same as for the Runge-Kutta scheme.  For
this new scheme to be monotone,  however, the time step has to satisfy
the following more restrictive condition
\begin{equation}
\left( \frac{u}{\Delta x}+\frac{v}{\Delta y} \right) \Delta t 
\le \frac{2}{\sqrt{2} + 1} \, .
\end{equation}

\begin{Prop}
For a uniform velocity field, if the initial data satisfy the integral
form  of  the solenoidality  constraint,  the {\it  Conservative-MUSCL
Scheme} for the  induction equation is identical to  a new, consistent
and  stable first  order  scheme  for the  advection  equation on  the
staggered grid.
\end{Prop}

At the expense  of a more restrictive constraint on  the time step, we
have obtain a new scheme  which is conservative for the predicted step,
in  the  sense  that   the  predicted  magnetic  field  satisfies  the
solenoidality  constraint. 

\subsection{High Order Schemes}

Extensions of the three above schemes (Runge-Kutta,
U-MUSCL and C-MUSCL) to second order are based on  a piecewise linear
reconstruction of 
each magnetic  field   component, using  ``magnetic  flux conserving''
interpolation at each cell interface. Following the MUSCL approach, one
can compute corner (or edge) centered interpolated quantities, using a
Taylor expansion both in time and space as follows, for $B_x$
\begin{eqnarray}
\nonumber
 \left<B_x\right>_{i+\half,j+\half}^{n+1/2,B}=\left<B_x\right>_{i+\half,j}^{n} 
+\left(\frac{\partial B_x}{\partial t}\right)_{i+\half,j}^{n}\frac{\Delta t}{2}
+\left(\frac{\partial B_x}{\partial y}\right)_{i+\half,j}^{n}\frac{\Delta y}{2}
\, ,\\
 \left<B_x\right>_{i+\half,j-\half}^{n+1/2,T}=\left<B_x\right>_{i+\half,j}^{n}  
+\left(\frac{\partial B_x}{\partial t}\right)_{i+\half,j}^{n}\frac{\Delta t}{2}
-\left(\frac{\partial B_x}{\partial y}\right)_{i+\half,j}^{n}\frac{\Delta y}{2} \, ,
\end{eqnarray}
and for $B_y$
\begin{eqnarray}
\nonumber
 \left< B_y \right>_{i+\half,j+\half} ^{n+1/2,L} = \left< B_y \right>_{i,j+\half} ^{n}  
+\left(  \frac{\partial B_y}{\partial t}  \right)_{i,j+\half} ^{n}   \frac{\Delta t}{2}
+\left(  \frac{\partial B_y}{\partial x}  \right)_{i,j+\half} ^{n}  \frac{\Delta x}{2}\, ,\\
 \left< B_y \right>_{i-\half,j+\half} ^{n+1/2,R} = \left< B_y \right>_{i,j+\half} ^{n}  
+\left(  \frac{\partial B_y}{\partial t}  \right)_{i,j+\half} ^{n}   \frac{\Delta t}{2}
-\left(  \frac{\partial B_y}{\partial x}  \right)_{i,j+\half} ^{n}  \frac{\Delta x}{2} \, .
\end{eqnarray}

In this way,   second-order, edge-centered components of  the magnetic
field can be used  in the  2D Riemann solver   to compute the  EMF and
update the solution  to time $t^{n+1}$.   Our three different  schemes
differ  in the way they implement  the terms $\partial B_x/\partial t$
and $\partial B_y/\partial  t$.  

Let us stress that  to recover second  order
accuracy in space,  one needs to perform  a predictor step which is also
second  order accurate in  space.   For  the C-MUSCL  scheme,  this is
already the case if one  uses exactly the  predictor step presented in
the last section.
For both the Runge-Kutta and  the U-MUSCL schemes, however, one needs to
use  a linear  reconstruction  of each  magnetic  field component  and
compute  the  EMF for  the  predictor step.  This  is  done using  the
following equations
\begin{eqnarray}
\nonumber
 \left<B_x\right>_{i+\half,j+\half}^{n,B}=\left<B_x\right>_{i+\half,j}^{n}  
+\left(\frac{\partial B_x}{\partial y}\right)_{i+\half,j}^{n}\frac{\Delta y}{2}\, ,\\
 \left< B_x \right>_{i+\half,j-\half} ^{n,T} = \left< B_x \right>_{i+\half,j} ^{n}  
-\left(  \frac{\partial B_x}{\partial y}  \right)_{i+\half,j} ^{n}  \frac{\Delta y}{2}\, ,\\
\nonumber
 \left< B_y \right>_{i+\half,j+\half} ^{n,L} = \left< B_y \right>_{i,j+\half} ^{n}  
+\left(  \frac{\partial B_y}{\partial x}  \right)_{i,j+\half} ^{n}  \frac{\Delta x}{2}\, ,\\
 \left< B_y \right>_{i-\half,j+\half} ^{n,R} = \left< B_y \right>_{i,j+\half} ^{n}  
-\left(  \frac{\partial B_y}{\partial x}  \right)_{i,j+\half} ^{n}  \frac{\Delta x}{2} \, .
\end{eqnarray}

These edge-centered components are then used to compute the EMF, using
(\ref{2DRPsol})     for    the    Runge-Kutta     method,    or
(\ref{1DRPsol_bottom_top})  and  (\ref{1DRPsol_left_right})  for
the  U-MUSCL scheme.  As usually  done in  higher order  finite volume
schemes, spatial derivatives are approximated using slope limiters, in
order to obtain positivity  preserving, non oscillatory solutions.  
For that  purpose we use a  standard slope  limiter (used  in  many fluid
dynamics codes), the Monotonized Central Limiter, which is given by
\begin{equation}
\left(  \frac{\partial B}{\partial x} \right)   = 
\rm{minmod}
\left( 
\frac{B_{i+1} - B_{i-1}  }{ 2\Delta x },
\rm{minmod}
\left( 
2\frac{B_{i+1} - B_i }{\Delta x}, 
2\frac{B_i - B_{i-1} }{\Delta x} 
\right)
\right) \, .
\label{moncen}
\end{equation}

Far  from  discontinuities,  this  slope  reduces  to  Fromm's  finite
difference approximation of the  spatial derivative. In this case, one
can show that,  for a uniform velocity field, all  3 schemes are again
strictly  equivalent  to their  second  order  parent  scheme for  the
advection equation on the staggered grid.

In non smooth parts of the flow, however, this is no longer true. Slope
limiting destroys the strict equivalence between the induction schemes
and  their  advection  counterparts.   One must  also  be  aware  that
traditional slope limiters, such as  the one we use here, are designed
for the advection equation  in finite-volume schemes. The monotonicity
of the solution for the induction equation is therefore not guaranteed.
Deriving slope limiters for the induction equation is beyond
the  scope of  this paper. We  have to  rely on  the numerical  tests
performed in the test section to assess the non oscillatory properties
of our schemes.

It is also apparent in (\ref{moncen}) that for both Runge-Kutta
and U-MUSCL  schemes, the computational stencil increases  by one cell
in  each   direction,  compared  to   the  first  order   scheme  (see
Fig.~\ref{figstencils}).   The second  order U-MUSCL  and  the C-MUSCL
schemes are therefore both  compact enough for our AMR implementation,
while the second order Runge-Kutta scheme is not.

\subsection{Conclusion}

We have derived in this section three numerical schemes for the solution
of the
induction equation  using the CT algorithm in  two-dimensions. All of
them are second order in space  and time. We have called these schemes
Runge-Kutta,  U-MUSCL and  C-MUSCL.  Only  the last  two  have compact
computational stencils, which makes them suitable for our tree-based AMR
implementation. More interestingly, we have  proven that, in case of a
uniform velocity  field, the U-MUSCL  scheme is strictly  identical to
Colella's Corner  Transport Upwind scheme for  the advection equation
on  the staggered grid. For  the C-MUSCL scheme,  we have shown
that  it  is  strictly  identical to  another  well-behaved  advection
scheme,  with however a  more restrictive  stability condition  on the
time step.  This shows that CT, when properly derived within Godunov's
framework,  has  advection  properties  similar  to  traditional
finite-volume schemes.

\section{A Constrained Transport AMR Scheme in three Dimensions}

In this section, we describe  our MUSCL-type schemes for the induction
equation  in  three   space  dimensions.   It is mostly a straightforward
generalization  of the  previous 2D  schemes,   we will however 
repeat each step  of the algorithm in order  to summarize our method,
and introduce the discussion of the AMR implementation.

\subsection{Definitions}

Let us generalize the schemes discussed in 2D in section~\ref{section2}
to 3D problems.
The three magnetic field components are discretized on a staggered
grid using a finite-surface representation
\begin{equation}
  \left< B_x \right>_{i+\half,j,k}^n = \frac{1}{\Delta y}\frac{1}{\Delta z}
  \int _{y_{i-1/2}} ^{y_{i+1/2}} \int _{z_{i-1/2}} ^{z_{i+1/2}} 
  B_x(x_{i+1/2},y,z,t^n)\, {\rm d}y\, {\rm d}z \, ,
  \label{bxdef3d}
\end{equation}
\begin{equation}
  \left< B_y \right>_{i,j+\half,k} ^n= \frac{1}{\Delta x}\frac{1}{\Delta z}
  \int _{x_{i-1/2}} ^{x_{i+1/2}} \int _{z_{i-1/2}} ^{z_{i+1/2}} 
  B_y(x,y_{i+1/2},z,t^n)\, {\rm d}x\, {\rm d}z \, ,
 \label{bydef3d}
\end{equation}
\begin{equation}
  \left< B_z \right>_{i,j,k+\half} ^n= \frac{1}{\Delta x}\frac{1}{\Delta y}
  \int _{x_{i-1/2}} ^{x_{i+1/2}} \int _{y_{i-1/2}} ^{y_{i+1/2}} 
  B_z(x,y,z_{i+1/2},t^n)\, {\rm d}x\, {\rm d}z \, .
 \label{bzdef3d}
\end{equation}
These three conservative  variables satisfy the divergence-free constraint
in integral form
\begin{eqnarray}
\nonumber
\frac{\left<B_x\right>_{i+\half,j,k}^n - \left<B_x\right>_{i-\half,j,k}^n}
{\Delta x} +
\frac{\left<B_y\right>_{i,j+\half,k}^n - \left<B_y\right>_{i,j-\half,k}^n }
{\Delta y} \\
+\frac{\left<B_z\right>_{i,j,k+\half}^n - \left<B_z\right>_{i,j,k-\half}^n }
{\Delta z} = 0 \, .
\label{divBint3d}
\end{eqnarray}

\subsection{Conservative update}

The  magnetic field  components are  updated from  time $t^n$  to time
$t^{n+1}$  using the induction  equation in  integral form,  which 
becomes (for $B_x$)
\begin{eqnarray}
\nonumber
\left< B_x \right>_{i+\half,j,k} ^{n+1} = 
\left< B_x \right>_{i+\half,j,k} ^{n} &+&
\frac{\Delta t}{\Delta y}
\left(  \left< E_z \right>_{i+\half,j+\half,k}^{n+\half} - 
        \left< E_z \right>_{i+\half,j-\half,k}^{n+\half} \right)\\
&-& \frac{\Delta t}{\Delta z}
\left(  \left< E_y \right>_{i+\half,j,k+\half}^{n+\half} - 
        \left< E_y \right>_{i+\half,j,k-\half}^{n+\half} \right) \, ,
\label{induction_num_x_3d}
\end{eqnarray}
see (\ref{induction_num_x}) for comparison.

Similar expressions can be derived for $B_y$ and $B_z$.
Here, $E_x$, $E_y$ and $E_z$ are time-averaged EMFs defined at
each cell edges.

\subsection{2D Riemann Solver}

Each  of these  EMFs components are  obtained as  the solution of  a 2D
Riemann  problem,   defined  by  4  initial  states   surrounding  the
corresponding edge.  The upwind  solution  of this 2D Riemann  problem
for $E_x$ is given by
\begin{eqnarray}
\nonumber
\left<E_x\right>_{i,j+\half,k+\half}^{n+\half} &=& 
 v \left({
 \left<B_z\right>_{i,j+\half,k+\half}^{n+\half,R}
+\left<B_z\right>_{i,j+\half,k+\half}^{n+\half,L}
}\right)/2  \\
\nonumber
&-&w \left({
 \left<B_y\right>_{i,j+\half,k+\half}^{n+\half,T}
+\left<B_y\right>_{i,j+\half,k+\half}^{n+\half,B}
}\right)/2  \\
\nonumber
&-&\left|v\right|
\left({
 \left<B_z\right>_{i,j+\half,k+\half}^{n+\half,R}
-\left<B_z\right>_{i,j+\half,k+\half}^{n+\half,L}
}\right)/2   \\
&+&\left|w\right|
\left({
 \left<B_y\right>_{i,j+\half,k+\half}^{n+\half,T}
-\left<B_y\right>_{i,j+\half,k+\half}^{n+\half,B}
}\right)/2  \, ,
\label{2DRPsol3dx}
\end{eqnarray}
Where the magnetic field components, labeled $n+1/2,R$; $n+1/2,L$;
$n+1/2,T$ and $n+1/2,B$ are the time-centered predicted states
interpolated at cell 
edges.
Similar expressions for $E_y$ and $E_z$ can be deduced by permutations.
%

\subsection{Predictor Step}

The predicted states of the magnetic field
are  obtained through a Taylor  expansion in
time and space. For $B_x$, this translates into

\begin{eqnarray}
\nonumber
 \left<B_x\right>_{i+\half,j+\half,k}^{n+1/2,B}=\left<B_x\right>_{i+\half,j,k}^{n} 
+\left(\frac{\partial B_x}{\partial t}\right)_{i+\half,j,k}^{n}\frac{\Delta t}{2}
+\left(\frac{\partial B_x}{\partial y}\right)_{i+\half,j,k}^{n}\frac{\Delta y}{2}
\, ,\\
\nonumber
 \left<B_x\right>_{i+\half,j-\half,k}^{n+1/2,T}=\left<B_x\right>_{i+\half,j,k}^{n}  
+\left(\frac{\partial B_x}{\partial t}\right)_{i+\half,j,k}^{n}\frac{\Delta t}{2}
-\left(\frac{\partial B_x}{\partial y}\right)_{i+\half,j,k}^{n}\frac{\Delta y}{2}
\, ,\\
\nonumber
 \left<B_x\right>_{i+\half,j,k+\half}^{n+1/2,B}=\left<B_x\right>_{i+\half,j,k}^{n} 
+\left(\frac{\partial B_x}{\partial t}\right)_{i+\half,j,k}^{n}\frac{\Delta t}{2}
+\left(\frac{\partial B_x}{\partial z}\right)_{i+\half,j,k}^{n}\frac{\Delta z}{2}
\, ,\\
 \left<B_x\right>_{i+\half,j,k-\half}^{n+1/2,T}=\left<B_x\right>_{i+\half,j,k}^{n} 
+\left(\frac{\partial B_x}{\partial t}\right)_{i+\half,j,k}^{n}\frac{\Delta t}{2}
-\left(\frac{\partial B_x}{\partial z}\right)_{i+\half,j,k}^{n}\frac{\Delta z}{2} \, .
\label{predictor3d}
\end{eqnarray}
Similar  expressions can  be written  for $B_y$  and $B_z$~.
The spatial  derivatives are computed  in each direction  using the
slope limiter function (\ref{moncen}).  Our three schemes differ
only  in the  way the  time derivative   is
estimated in the  above  expansion.

\subsubsection{Runge-Kutta Scheme}

The Runge-Kutta  predictor step is  equivalent to the  corrector step,
except for  the time derivative in  (\ref{predictor3d}). We use
spatial derivatives to  define edge-centered magnetic field components
and the 2D Riemann solver  to define the edge-centered EMF components.
This unique EMF vector, defined at  time $t^n$, is finally used in the
conservative  formula   (\ref{induction_num_x_3d})  to  obtain  a
finite   difference   approximation   of   the  time   derivative   in
(\ref{predictor3d}).  For a uniform  velocity field,  the first
order  scheme is  again identical  to the  Runge-Kutta scheme  for the
advection equation on the staggered grid. For the second order scheme,
this is only true in smooth regions of the solution.

\subsubsection{U-MUSCL Scheme}

For the U-MUSCL scheme, the EMF used to compute the predicted states is
not  uniquely defined at  each edge  anymore,  so  that the  predicted
magnetic field does not  satisfy  the divergence-free  constraint.  In
fact, we  compute at each cell  edge 4  EMF components, specialized to
each  face-centered magnetic field component.  By solving a 1D Riemann
problem at each faces,  we perform  a  proper upwinding in the  normal
direction.    The  input  states of    these   1D Riemann problem  are
reconstructed magnetic  field  components at  cell edges  using  slope
limiters. Note that  for a  uniform velocity  field,  this first order
scheme is not equivalent anymore to the CTU scheme in 3D. 

\subsubsection{C-MUSCL Scheme}

Like the Runge-Kutta method, the C-MUSCL scheme involves one single
EMF  vector to compute  the time-derivative  in the  Taylor expansion,
therefore preserving the solenoidal property on the predicted step.
This  EMF  is computed  using  the  average of  the
face-centered magnetic  field components, as  in (\ref{CMUSCL}).  It
 does not  involve any limited  slope computations,
but still retains second order accuracy in space.  As explained in the
previous section,  the cost is  a more  restrictive time-step
stability   condition.  For a uniform  velocity field the scheme is identical
to the new advection scheme on the 3D staggered grid discussed in 
section~\ref{sectCMUSCL}.

\subsubsection{Merits of the Various Schemes}

We compare, in this section, the different advantages and drawbacks of 
each of the above described methods. The corrector step is
the same for each cases.

The Runge-Kutta scheme is the most natural scheme to write. However, 
it will prove to be very expensive for MHD, since it requires 
a 2D Riemann solver in the predictor step. Moreover, it has a
restrictive Courant condition and its stencil is too large to be 
implemented in the AMR  implementation, which is not the case of the two 
other schemes.

The U-MUSCL scheme has better stability properties, the time step is less 
restrictive. It is also expected to be more efficient in MHD
applications, since one 1D Riemann problem only is required 
in the predictor step. Note however that its rigorous 3D extension is 
problematic and requires further investigation.

Unlike the U-MUSCL scheme, for which the non-conservation of the
solenoidality condition in the predictor step may cause problems in 
some cases, the C-MUSCL scheme is conservative. No Riemann solver 
is needed in the predictor step, which should make it very efficient 
for MHD (\cite{Fromang06}). 
But these advantages are obtained at the cost of a smaller timestep 
than the U-MUSCL scheme.

\subsection{AMR Implementation}

We have included both of the compact schemes (U-MUSCL and C-MUSCL)
in the RAMSES  code. It is a tree-based  AMR  code originally designed
for  astrophysical    fluid dynamics   \citep{Teyssier02}.   The  data
structure is a ``Fully Threaded Tree'' \citep{Khokhlov98}. The
grid  is divided into  groups of 8  cells, called ``octs'', that share
the same parent cell.  Each oct has access  to its parent cell address
in  memory,  but  also to  neighboring  parent cells.  When  a cell is
refined, it is called a ``split'' cell, while in the opposite case, it
is called a ``leaf'' cell.  The computational domain is always defined
as the unit  cube, which corresponds in  our terminology to the  first
level of  refinement  in the  hierarchy $\ell =  1$. The  grid is then
recursively refined up to the minimum level of refinement $\ell_{min}$,
in order to build  the  coarse grid.   This coarse  grid is the   base
Cartesian grid,  covering the  whole computational domain,  from which
adaptive refinement can proceed.  This base grid is eventually refined
further up to some maximum level of refinement $\ell_{max}$, according
to some user defined refinement criterion.

When $\ell_{max}=\ell_{min}$, the  computational grid is a traditional
Cartesian grid, for which the previous induction schemes apply without
any  modification.   When refined  cells  are  created, however,  some
issues specific to AMR must be addressed.

\subsubsection{Divergence-free Prolongation Operator}

When a cell is refined, eight new cells (i.e.  a new ``oct'') are created
for which  new magnetic field components are  needed.  More precisely,
each of the six faces of the parent  cell are split into 4  new fine faces.
Three new faces, at the center of the parent cell, are also split into
four new children  faces.  The resulting magnetic
field  components,   fine  or  coarse, need to  satisfy  the
divergence-free constraint in integral form.

This critical  step, usually called  in the multigrid  terminology the
Prolongation  Operator,  has   been  solved  by  \cite{Balsara01}  and
\cite{Toth02} in the CT framework.  We recommend
both of these articles for  a detailed description  of the method.
The idea is  to used slope limiters to  interpolate the magnetic field
component inside each parent face,  in a flux-conserving way, and then
to use a  3D reconstruction, which is divergence-free  in a local sense
inside the  whole cell  volume, in order  to compute the  new magnetic
field components for each central  children faces. In our case, 
the same slope limiter as in the Godunov scheme (\ref{moncen}) has been
used.

This prolongation  operator is used  to estimate the magnetic  field in
newly refined cells,  but also to define a  temporary ``buffer zone'',
two ``ghost cells''  wide, that set the proper  boundary for fine cells
at a coarse-fine level boundary. This  is the main reason why 
a compact stencil is needed
for the underlying Godunov scheme.

\subsubsection{Magnetic Flux Corrections}

The other  important step is to  define the reverse  operation, when a
split cell is de-refined, and becomes a leaf cell again. This operation
is  usually  called the  Restriction Operator
in  the  multigrid  terminology. 
The solenoidality constraint needs again to be satisfied,
which translates into conserving the magnetic flux. The magnetic field
component in the  coarse face is just the arithmetic  average of the 4
fine face values. This is  reminiscent of the ``flux correction step''
of   AMR  implementations   for  Euler   
systems~\citep{Berger84,Berger89,Teyssier02}.

\subsubsection{EMF Corrections}

The ``EMF correction  step''     is more specific to    the  induction
equation. For a coarse face which is adjacent, in  any direction, to a
refined face, the coarse EMF in the  conservative update of  the
solution needs to be replaced by the arithmetic average of the two fine EMF
vectors.  This    guarantees   that  the    magnetic field     remains
divergence-free, even at coarse-fine boundaries.

\subsection{Physical resistivity }

We  have now  completely described  our AMR implementation for  the induction
equation.   It can  be used  as such, without  explicitly including
physical  resistivity, to  investigate  fast-dynamo action  associated
with a  given flow.   The resulting integration  is  stable and
produce an exponentially growing field  very similar to what we expect
in dynamo theory.  However,  resistivity (and thus reconnection), which
is necessary  to identify  a growing eigenmode, is solely due  to the
underlying  numerical scheme.  This  numerical resistivity  is usually
non-uniform  in  time  and  space, anisotropic  and  non-linear.   The
mathematical properties  of the resulting eigenmodes  are unclear, and
the results  usually depend on  the  mesh resolution.   Instead, we
have  chosen  to  explicitly  introduce a physical  resistivity  in  the
induction equation, see (\ref{magnetic induction eq adim}), in order to allow 
a proper identification of the eigenmode.

The amplitude of  the resistive term is here  controlled by the inverse
of  the magnetic  Reynolds number  $Rm=UL/\eta$. 
We shall concentrate on large magnetic Reynolds
numbers (i.e. the fast dynamo  limit).  It may, at first, seem strange
to introduce  this term when  the Godunov approach has  precisely been
introduced  to  ensure   numerical  stability  and  reduced  numerical
diffusion.  In  fact, because  of the very  nature of the  fast dynamo
solution, the effect  of physical resistivity will be  limited to very
localized regions.  Its effect will  therefore be limited to  the very
fine AMR  cells and the  stabilizing property of the  Godunov approach
will be essential for the coarser cells.

Physical diffusivity is introduced in  our scheme using
the  operator splitting  technique. After  the induction  equation has
been  advanced to  the next  time coordinate  $t^{n+1}$  with solution
$\B^*$, we solve for the diffusive source term, using the following equation
\begin{equation} 
\frac{\B^{n+1}-\B^*}{\Delta t} = \eta \vnabla \times \j^{n+1}
~~~\rm{where}~~~
\j^{n+1} = \nabla \times \B^{n+1} \,,
\label{magnetic diffusion eq} 
\end{equation} 
where $\j$  is the  current.  It is  defined at cell  edges.
For example, the finite
difference approximation for $j_x$ ($j_y$ and $j_z$ are not shown) is written as
\begin{eqnarray}
 (j_x)_{i,j+\half,k+\half} = 
 \frac{
 \left<B_z\right>_{i,j,k+\half}^{n+1}
-\left<B_z\right>_{i,j,k+\half}^{n+1}
}{\Delta y}
- \frac{
 \left<B_y\right>_{i,j+\half,k}^{n+1}
-\left<B_y\right>_{i,j+\half,k}^{n+1}
}{\Delta z} \, .
\label{current_jx}
\end{eqnarray}

Considering the current as the  analog of the EMF, all the ingredients
of the previous sections can  be applied to design a conservative  AMR
implementation to solve for the diffusion source term. We use for that
purpose 
a fully implicit time discretization, in order for the time step to be
limited only by the  induction scheme Courant stability condition. The
resulting linear system is solved iteratively using the Jacobi method. Note
that in the problems we address in this paper, only a few iterations were
necessary to reach $10^{-3}$ accuracy. 

\section{Tests and Application to Kinematic Dynamos}

In this section, we test our  various schemes using the advection of a
magnetic field loop in 2D. We  conclude that the three Godunov schemes we
described for
the  induction equation have very  good and  similar performances.
The U-MUSCL scheme seems to be  slightly better than the other two. We
also  test  the AMR  implementation,  showing that  the  results  are  almost
indistinguishable from  the reference Cartesian run.  We  will then use
this code  to compute the  evolution of two well-studied  dynamo flows:
the Ponomarenko dynamo and the ABC flow. This will serve as a final 
integrated test of our scheme.

\subsection{Magnetic Loop Advection}

\begin{figure}
\centerline{\epsfxsize 15cm \epsffile{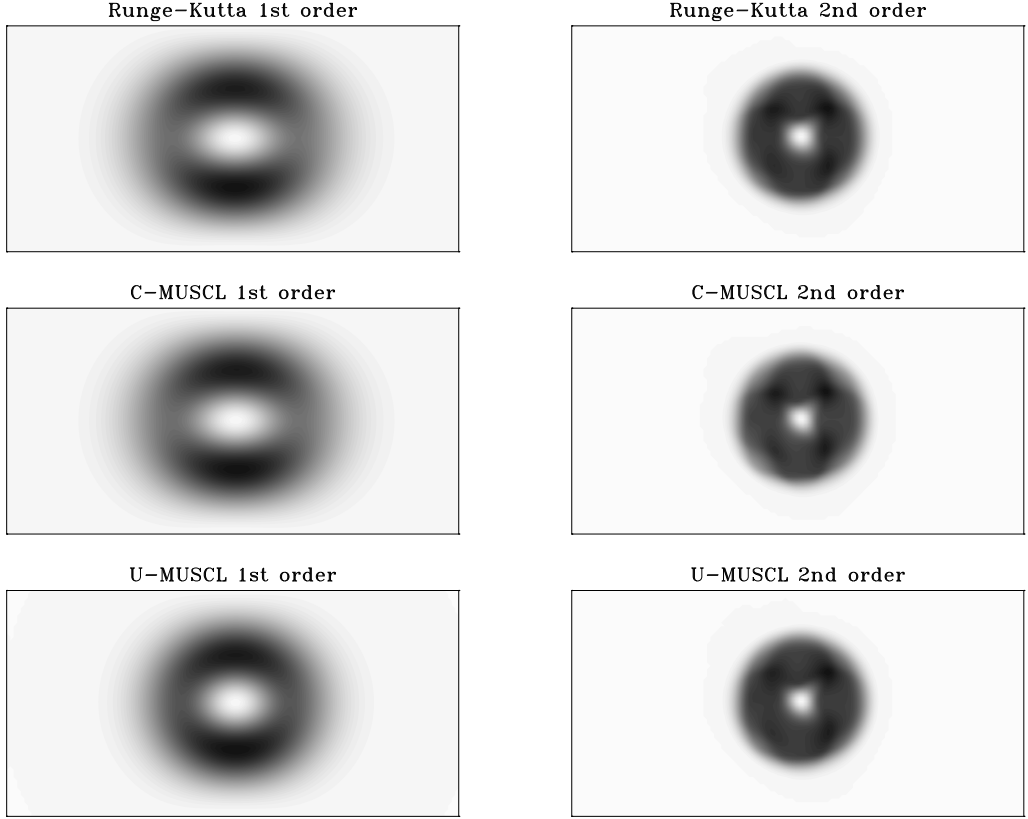}}
\caption{Magnetic  loop  advection  test  for a  Cartesian  grid  with
$n_x=128$ and  $n_y=64$: each  panel shows a  gray-scale image  of the
magnetic  energy ($B_x^2+B_y^2$)  at time  $t=2$. The  scheme  used to
compute each  image is provided  in the title of each panel. Second-order  
schemes give
very similar  results, while the  first order U-MUSCL  scheme performs
slightly better than the two other first order schemes.}
\label{figloopcomp}
\end{figure}

\begin{figure}
\epsfxsize 15cm
\epsffile{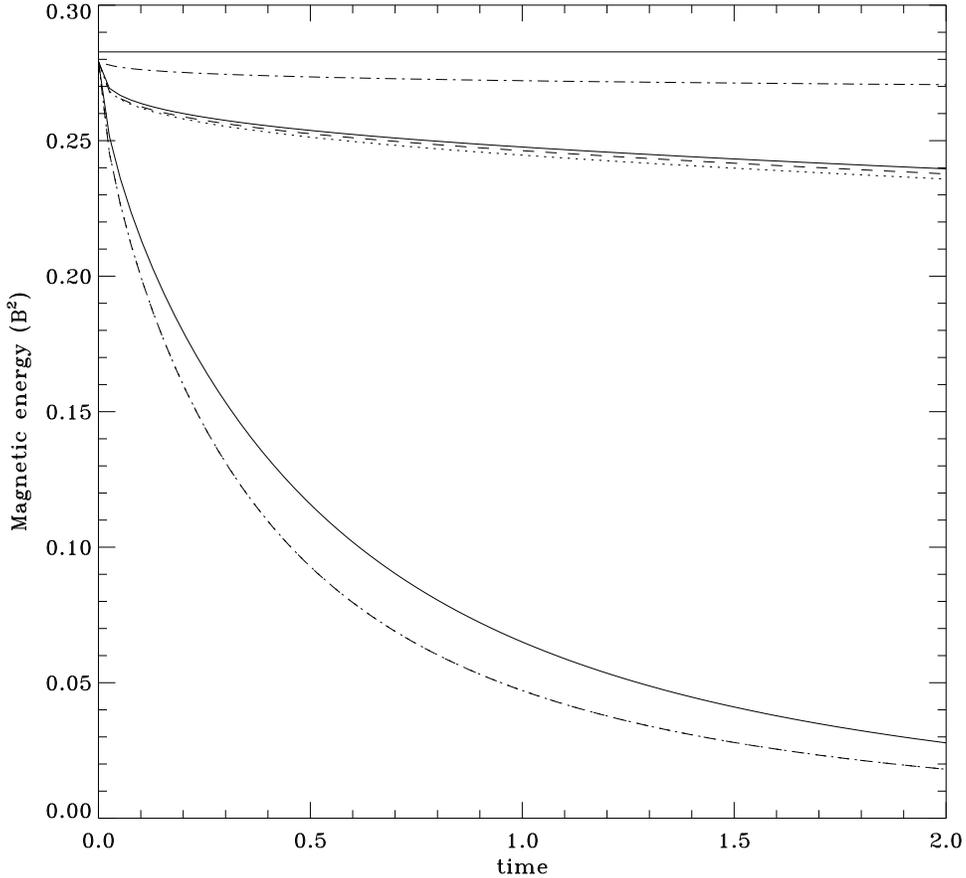}
\caption{Magnetic  energy as  a function  of time  for the  field loop
advection test.   The upper solid line  is the solution  for perfect
advection.   The  lower  lines   are  for  the  first  order  schemes:
Runge-Kutta (dotted  line), C-MUSCL  (dashed line) and  U-MUSCL (solid
line).     Runge-Kutta   and    C-MUSCL    results   are
indistinguishable in this case. The 3 intermediate lines correspond to 
second order schemes and use the same line convention. The dot-dashed
lines is the AMR result obtained with U-MUSCL and using $\ell_{min}=3$ and
$\ell_{max}=9$.} 
\label{figloopenergy}
\end{figure}

Let us first focus our attention  on a simple test of pure
advection which was  recently proposed by  \cite{Gardiner05} to
investigate the  advection properties of their CT  scheme. It consists
in  the advection of  a magnetic  field loop  with a  uniform velocity
field. It is of particular relevance in our case, since we are dealing
with kinematic induction problems. The computational domain is defined
by $-1 < x < 1$ and $-0.5 < y < 0.5$. The boundary conditions are periodic.
The flow velocity is set to $u=2$, $v=1$ and $w=0$.

The initial magnetic field is such that $B_z=0$, while 
$B_x$ and $B_y$ are defined using the z-component of the potential
vector $\A$ (with $\B= \vnabla \times \A$), as an axisymmetric function
of the form
\begin{equation}
A_z = \left\{
  \begin{array}{cc}
    R-r & \rm{for~} r < R \, ,\\
    0 & \rm{otherwise} \, ,
  \end{array}
\right. 
\end{equation}
with $R=0.3$ and $r=\sqrt{x^2+y^2}$.  The exact amplitude of
the  magnetic  field is  arbitrary,  since  we  are solving  a  linear
equation, we used $B=1$.  In the following, we use exactly the same
resolution as \cite{Gardiner05}.

We  perform  the  numerical integration  of  the  induction
equation up to time $t=2$ with a Courant factor see (\ref{cfl}) is equal 
to $0.8$,  for which the magnetic loop has evolved 
twice across the computing box. Our first  set of  runs use  a regular
Cartesian grid with  $N_x=128$ and $N_y=64$.  We test  the three different
schemes,  to first  order (slope  limiters were  set to  zero)  and to
second order. The aim here is to estimate  the  numerical diffusion
of our  various schemes.   

Figure~\ref{figloopcomp}  shows  gray-scale  images  of  the  magnetic
energy $B_x^2+B_y^2$ for the six runs.  Maximum field dissipation occurs
at the center and boundaries of  the loop where the current density is
initially  singular.   Second  order  schemes all  give  very  similar
results.  At first order,  the U-MUSCL scheme performs slightly better
than the other  two, with a more isotropic  pattern.  To estimate more
quantitatively   the   numerical  diffusion,   we   have  plotted   in
Figure~\ref{figloopenergy}   the   total   magnetic  energy   in   the
computational box as a function of time.  Perfect advection would have
given a constant value of  $E_{tot}=\pi R^2$.  As expected, first order
schemes are much more diffusive than the second order ones.  
All the latter  give almost identical results, Runge-Kutta  being the most
diffusive, followed by  C-MUSCL and then U-MUSCL. At  first order, the
U-MUSCL scheme also appears less diffusive than the two other schemes.

\begin{figure}
\epsfxsize 15cm
\epsffile{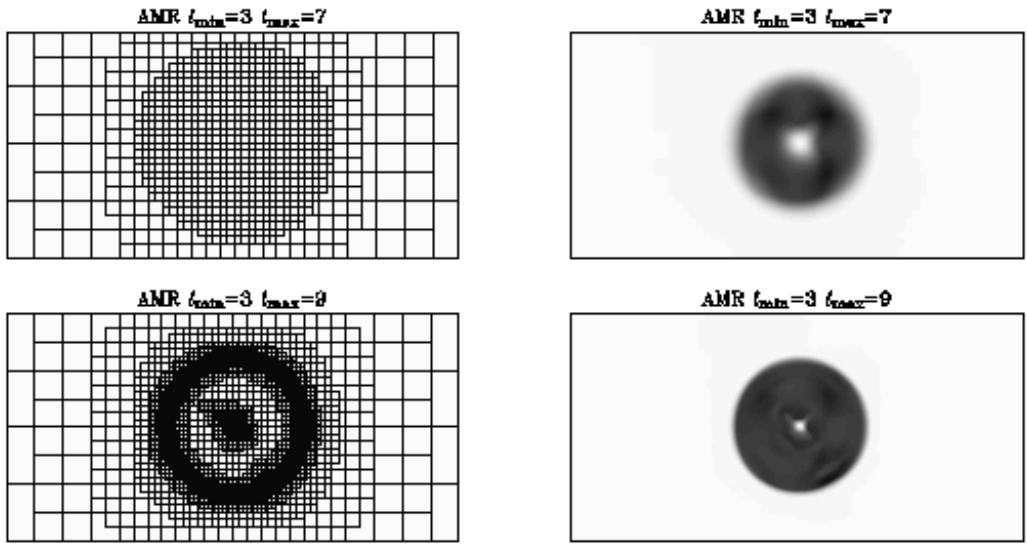}
\caption{Magnetic  loop advection  test: AMR  result with  the U-MUSCL
scheme.  The  two upper  plots are for  $\ell_{max}=7$, while  the two
lower plots are for  $\ell_{max}=9$. The right panels show gray-scale
images of  the magnetic  energy, while the  left panels show  the AMR
grid (only ``oct''  boundaries are shown for clarity,  but each oct is
in fact subdivided into 4 children cells).}
\label{figloopamr}
\end{figure}

We now present  the results obtained with our AMR implementation using the U-MUSCL
scheme (C-MUSCL giving almost identical results). We start with a base
Cartesian   grid   with   $N_x=8$   and  $N_y=4$,   corresponding   to
$\ell_{min}=3$.    It is then adaptively  refined up  to
$\ell_{max}$, using the following refinement criterion on the magnetic
energy $E = B_x^2 + B_y^2$

\begin{equation}
\frac{\max{ \left( \left| \Delta_x E \right|, \left| \Delta_y E \right|
\right) } }{E+0.01}> 0.05 \, .
\end{equation}

With this  criterion, each cell for  which the change of local magnetic energy exceeds 5\% of
the local magnetic  energy is refined.   The first test is done  with
$\ell_{max}=7$, in order to  reach the same  spatial resolution as the
previous simulations with  a $128\times64$ Cartesian grid.  The magnetic
energy map at $t=2$ is shown in Figure~\ref{figloopamr}, together with
a line plot  showing the  corresponding  AMR grid. In this  last plot,
only ``oct'' boundaries are  shown for clarity   (each oct is in  fact
subdivided into four children    cells).   We conclude that  the   AMR
results are indistinguishable from the equivalent resolution Cartesian
run, but the computational  cost\footnote{The actual computing time 
is in our case directly proportionnal to the number of active cells. }
is  lower:  at time $t=2$, the  total
number of leaf cells in the AMR tree is $3149$. This is to be compared
with the number of cells in  a Cartesian grid  equivalent to the finer
resolution which would be $128\times64=8129$.

In order to illustrate more convincingly the interest  of using an AMR
grid in  this case, we   have performed the  same simulation  with now
$\ell_{max}=9$. The magnetic energy map and the corresponding AMR grid
are shown in  Figure~\ref{figloopamr}.  Refinements are  now much more
localized   at the  center  and    boundaries  of the  magnetic  loop.
Numerical  diffusion  has   dramatically   decreased,   as  shown   on
Figure~\ref{figloopenergy},   where the   time  history  of  the total
magnetic energy is   plotted. The agreement  with  the ideal case  has
improved substantially.  The total  number of  cells  at $t=2$ is  now
$16433.$   This is only   a  factor of   2 greater  than  the previous
Cartesian  runs, but  a  factor of   8  lower than the  Cartesian grid
equivalent to the finer resolution $512\times256=131072$.

\subsection{The Ponomarenko Dynamo}

One of the simplest known dynamo flows, and the one we will start our
investigation with, is the Ponomarenko dynamo \citep{Pono73}. The geometry 
of the flow is remarkably simple. In cylindrical polar coordinates
$(s,\phi,z)$, it is
\begin{equation}
\v = \left\{
\begin{array}{ll}
(0,s \Omega, u_z)\mbox{\ \hskip 1cm\ }& \mbox{for\ } s \leq s_0 \, , \\
\zero& \mbox{for\ } s > s_0 \, .
\end{array}
\right.
\end{equation}
This flow features an abrupt discontinuity across
the   cylinder at  $s=s_0$, such discontinuity yields an intricate
behavior in the  limit  $\Rm \rightarrow  \infty$. The  growth rate remains
constant in this limit, but the flow does not qualify as a proper fast
dynamo,   for  the critical eigenmode keeps    changing with $\Rm$ (see
\citeauthor{gilbert}, \citeyear{gilbert}). Variants of this
flow,  known  as  ``smoothed  Ponomarenko  flows''  introduce  a
typical length   scale over  which the flow   vanishes,  and  can help
circumvent   this  difficulty   \citep{Gilbert88}.   We  will  however
consider    here  the  original    Ponomarenko  flow   with  an abrupt
discontinuity. Since the  flow is discontinuous, an  explicit physical
resistivity  (associated with  a finite value   of the Reynolds number
$\Rm$) is essential in setting   the typical lengthscale  of the
magnetic field ($\ell \sim \Rm ^{-1/2}$).

As with most  dynamo problems, numerical  resolution is classically
achieved  using   spectral  expansions  (e.g.    \citeauthor{gilbert},
\citeyear{gilbert}).   We use here our   numerical approach to validate 
our scheme as well as to test the properties of the AMR implementation and its
ability to  deal with a  discontinuous input flow.  
Because of the cylindrical nature of the  flow, it is natural to think
of  adapting  the scheme  to   this  system of  coordinates.  We  have
therefore written a cylindrical version of our algorithm (note however
that AMR has not been implemented  in  this version of  the
code).    The discontinuity at $s=s_0$    correspond exactly to a cell
boundary. It is important to appreciate that there
is no  flow along the  $s$  direction with this approach. 
This implies  that  numerical
diffusion  vanishes in  this  direction.  It  is only  nonzero  in the
$\phi$ and $z$   directions.  This emphasizes  the importance  of  
physical resistivity to obtain meaningful results.

In  most practical work, sharp structures  in the flow can occur which
are  not necessarily aligned with the  grid  (see for example the next
application). We will therefore solve  this same dynamo problem  using
also a Cartesian grid.  A very large  resolution is needed in order to
reach  a fine discretisation of  the cylinder at $s=s_0$ (around which
the   field  is localized   over     a  lengthscale $\ell  \sim    \Rm
^{-1/2}$). This will be achieved using our AMR approach.

The Ponomarenko flow can  be investigated analytically \citep{Pono73}.
Such an  analysis reveals  that  an exponentially  growing solution in
time can  be obtained  for $\Rm =   U s_0 /   \eta \geq  \Rm_c  \simeq
17.7$~(where  $U=\sqrt{\Omega^2   s_0^2+u_z  ^2}$).  This  is obtained
using a spectral expansion  of the variables in $z$  and $\phi$ of the
form $\exp(i  m \phi + i  k z)$.  The most  unstable mode (at  $\Rm =
\Rm_c$) corresponds to  $u_z= 1.3 \,  \Omega  s_0$, $m=1$ and  $k_c \,
s_0=0.39$.  For larger  magnetic   Reynolds number, other  modes
become unstable.

\begin{figure}
\epsfxsize \textwidth
\epsffile{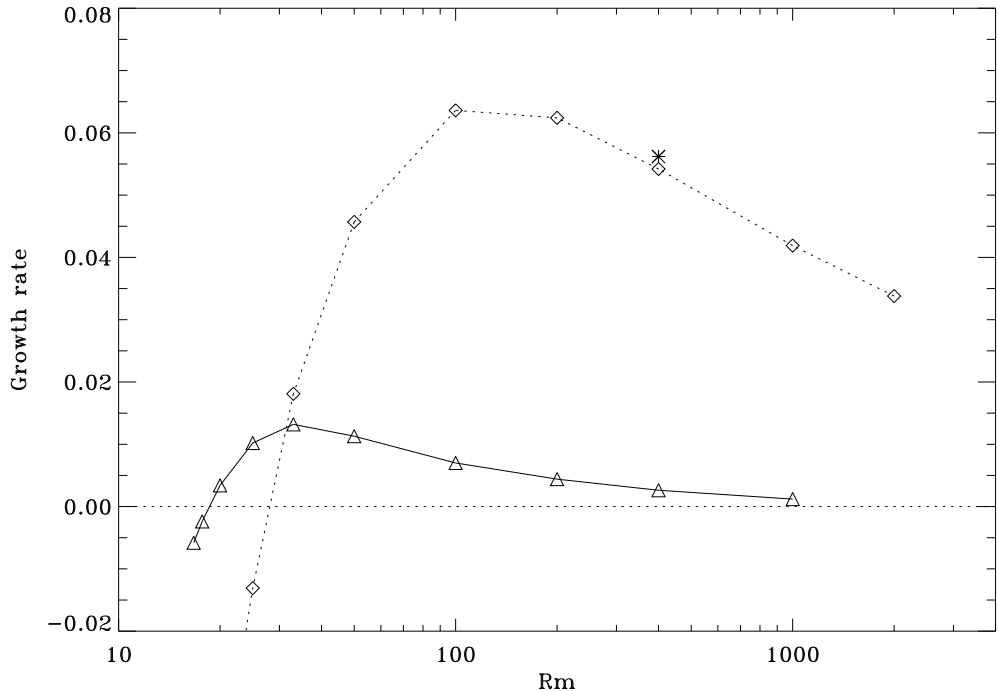}
\caption{Growth rate for the Ponomarenko  dynamo as a function of  the
magnetic Reynolds number.  The solid  curve  corresponds to the  first
unstable mode,  and the    dotted  line to its    harmonic  $k=2\times
k_c$~.   For both  modes,  the growth  rate   first increases and then
decreases with $\Rm$ (as expected  from analytical linear theory).  As
the  Reynolds number   increases, a transition occurs  from  $k_c$  to
$2\times  k_c$~.   The star symbol  $\Rm=400$   corresponds to the AMR
simulation. }
\label{figponorate}
\end{figure}

Using  the cylindrical   version  of  our  code,  we  have numerically
calculated the magnetic energy growth rates   for the Ponomarenko
flow   for a  large  range of  magnetic  Reynolds  numbers, going from
$Rm=16.7$ to $Rm=2000$. 

We use $s_0$ as unit of length, thus senting $s_0\equiv 1$.
The grid extends  from $0.2$  to   $3.5$ in radius and  the  azimuthal
coordinate cover the full $2\pi$ range.  The resolution of the grid is
$(N_r,N_{\phi},N_z)=(64,50,64)$.
For the vertical  extent of the computational domain $L_{\rm box}$,  we
consider two different
cases: {\bf case  I}, for which  $L_{\rm box}=2\pi/k_g$, with $k_g$ being
$0.39$ and {\bf case  II} for which  $k_g = 0.78$.  Let us recall that
the classical numerical approach for this  problem relies on a Fourier
expansion in $z$. In this case, a  single mode $k$  is retained in $z$
to enlighten the numerical procedure, the optimal value of $k_c$ being
obtained after optimization.   Our numerical approach  does not  allow
this  sort   of mode selection.      Instead,  we can   only fix   the
$z$-periodicity of the  computational box. In  case I,  $L_{\rm box}$ 
was chosen to match  the wavelength  of  the most unstable mode.  However,
harmonics of   the critical mode, being  unstable for large
Reynolds numbers, can   also  develop  in the
computational box (as can  be seen for example in  the figure  $6.4$ of
\citeauthor{Plunian}, \citeyear{Plunian}). This is a known issue, which only
occurs here because the 
calculation is not restricted to a single mode in~$z$.  

The transition  from the first unstable mode  to a higher mode  in $z$
occurs  for   Reynolds numbers twice  critical.  We  have been able to
follow the first unstable   mode to Reynolds  number larger   than the
transition to $k=2\times k_c$  by carefully  selecting the
initial condition (and restricting  to short enough time integrations). We
have  also turned our   attention to  the $k=2\times k_c$  instability
below  the transition by   studying  a computational   box of half  the
standard  size in  the    $z$-direction.  The  resulting    diagram is
presented in figure~\ref{figponorate}~.

When $Rm=16.7$, the  growth rate $\sigma$ of  the magnetic energy  was
found to be   negative, as expected. For  $Rm  \in [17.7,20]$ $\sigma$
becomes    positive in case~I   and   the   eigenmode corresponds   to
$k=k_c$~. When $Rm=20$, it is characterized by $m=1$, $k=k_c=0.39$ and
$\sigma=3.4  \times 10^{-3}$. This   is in very  good  agreement with
linear theory \citep{Pono73}. The growth rate obtained for larger $Rm$
is represented by the solid line on figure~\ref{figponorate}.

In case~II,  we use  a computational  domain  with half the  vertical
extend  of case~I. The  growing mode has   different properties. It is
characterized by $m=1$ and $k=2\times  k_c=0.78$. Its growth rate as a
function of $Rm$ is shown on figure~\ref{figponorate} using the dotted
line. The  transition  between both modes  is  clear near  $\Rm \simeq
30$. Unless the initial conditions are carefully chosen and the time 
integration is short enough, the mode $k=2
\times   k_c$ will  overcome the  first   critical mode for $\Rm > 30$~.

\begin{figure}
\centerline{
\epsfxsize 9cm
\epsffile{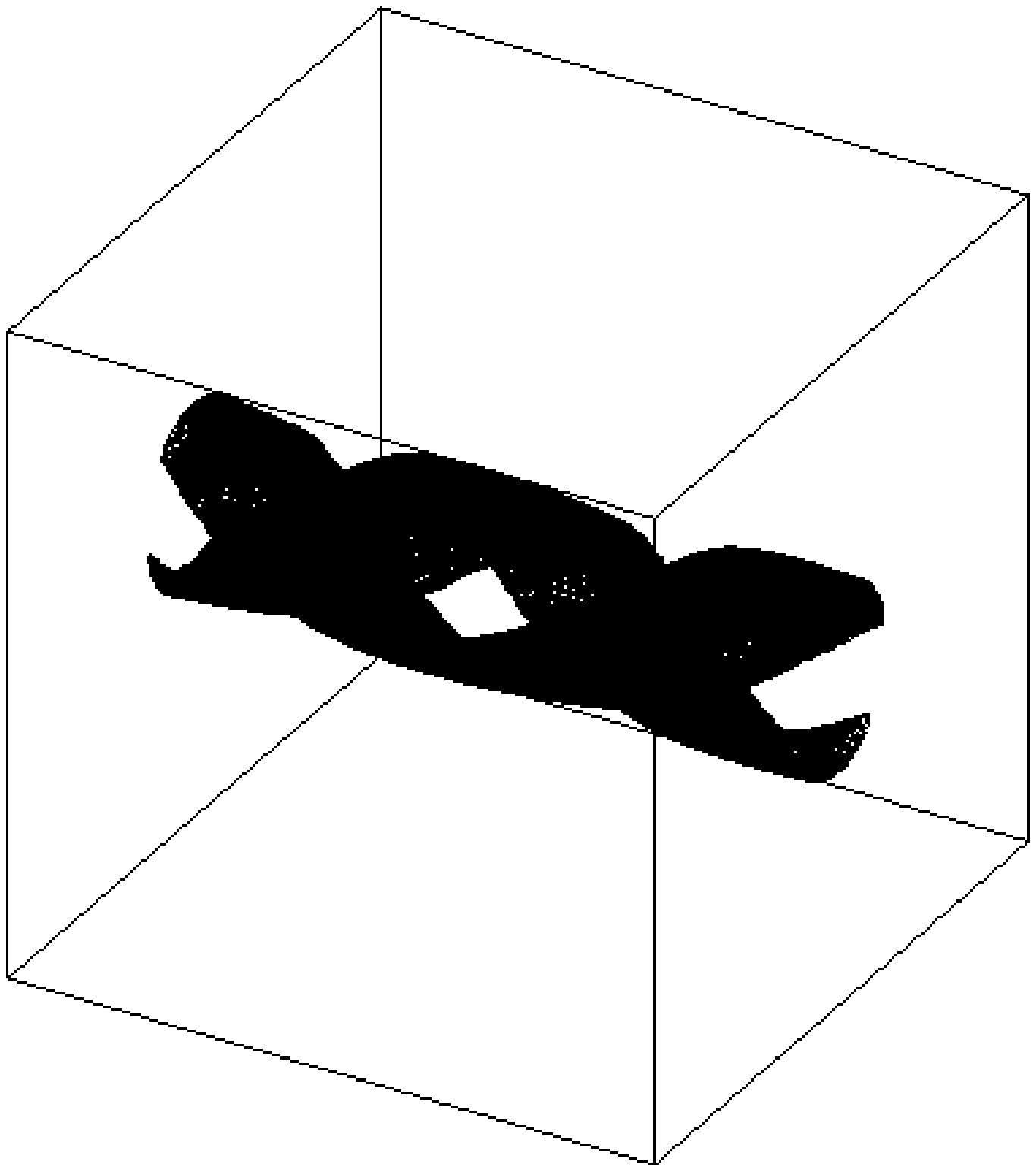}
\hskip -2cm
\epsfxsize 9cm
\epsffile{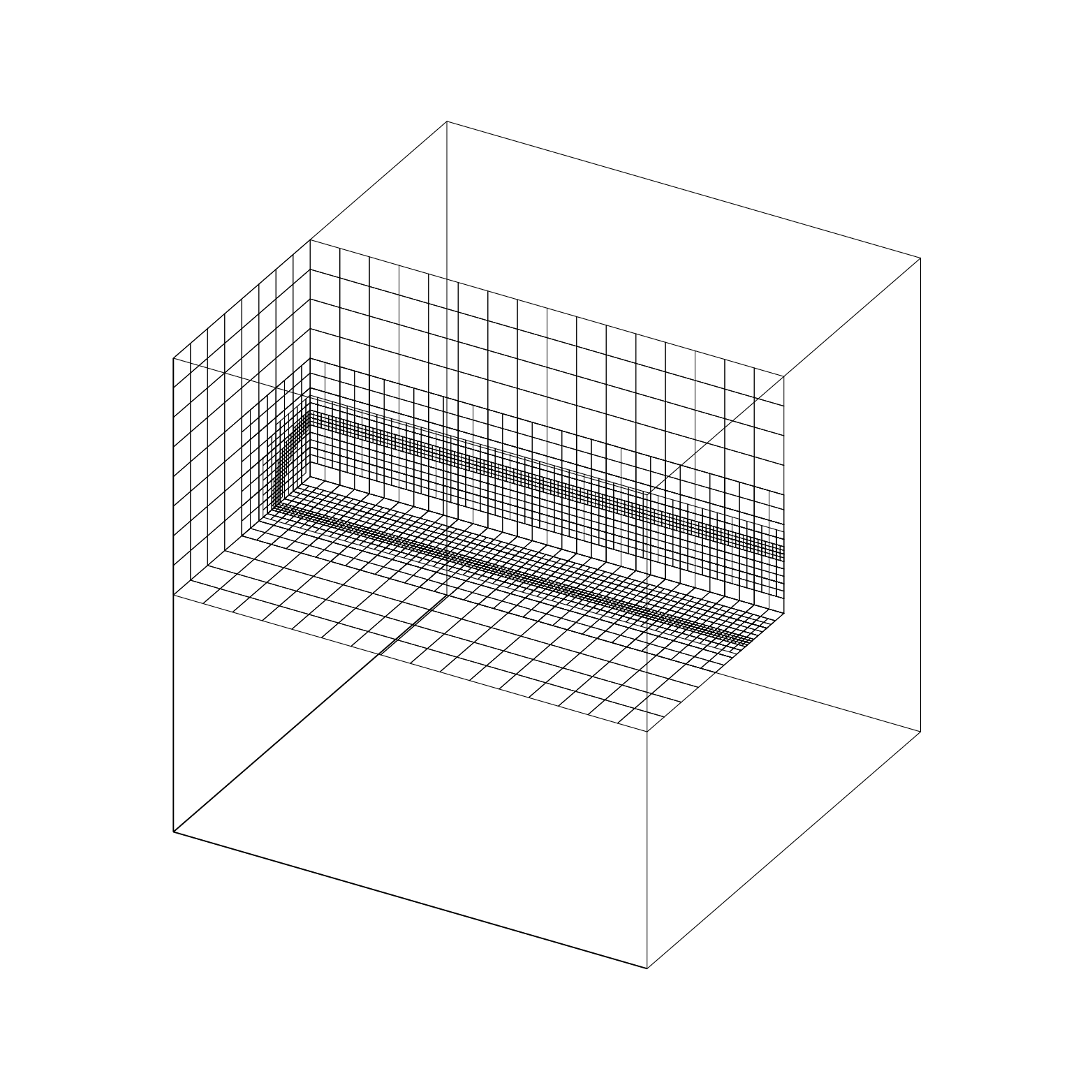}}
\caption{Ponomarenko dynamo  with  $\Rm=400$. Left panel:  surface  of
isovalue $B^2/2=10^6$ for the magnetic energy density at time $t=200$.
Right   panel: mesh geometry  (for clarity, only ``octs'' boundaries are
displayed here).} 
\label{figponoamr}
\end{figure}

In order to validate the AMR implementation, we  have also performed
simulations on a Cartesian grid
with  $\Rm=400$.  The   size   of the  box  is
$L_{\rm box}=2\pi/0.78$, similar to   case~II described above.   For  this
run, we took $\ell_{\rm min}=5$ and  $\ell_{\rm max}=8$, which has yield a maximum
of $751360$ cells on  the grid (this is  a factor of $22$ smaller than
the  number  of  cells  of  a $256^3$   uniform grid).  The refinement
strategy  was based  on the magnitude  of  the velocity gradient.  The
growth rate  of the magnetic energy  in  this case was measured  to be
$\sigma_{\rm AMR}=0.0562$       (see   the       star   represented     in
figure~\ref{figponorate}).   This is in  very  good agreement with the
value $\sigma=0.0542$  obtained with  the  cylindrical version  of the
code for the same parameter set.  

The structure of the growing eigenmode in  this simulation is illustrated
in figure~\ref{figponoamr}. The  left panel represents surfaces of isovalue
of the magnetic energy density $B^2/2$ at  $t=200$ while the structure
of the AMR grid is illustrated on the right panel.  The grid
is   only refined at the  sharp  boundary between   the inner rotating
cylinder and     the  outer   motionless medium.     This   simulation
demonstrates both the ability of  the scheme  to
simulate the Ponomarenko dynamo using a Cartesian grid
and the possibility to handle discontinuities in
the flow which  are not aligned  with the grid.

\subsection{The ABC Dynamo}

We now consider another dynamo flow, known as the ABC-flow (for
Arnold-Beltrami-Childress). It is defined by a periodic flow
\begin{equation}
\u = A \, (0,\sin x,\cos x) + B \, (\cos y,0,\sin y) 
+ C \, (\sin z,\cos z,0) \, .
\end{equation}
We limit our attention here to the classical case of $(A:B:C)=(1:1:1)$.
Let us  stress that this test is fully 3D and requires a significant
computational effort.

\begin{figure}
\epsfxsize \textwidth
\epsffile{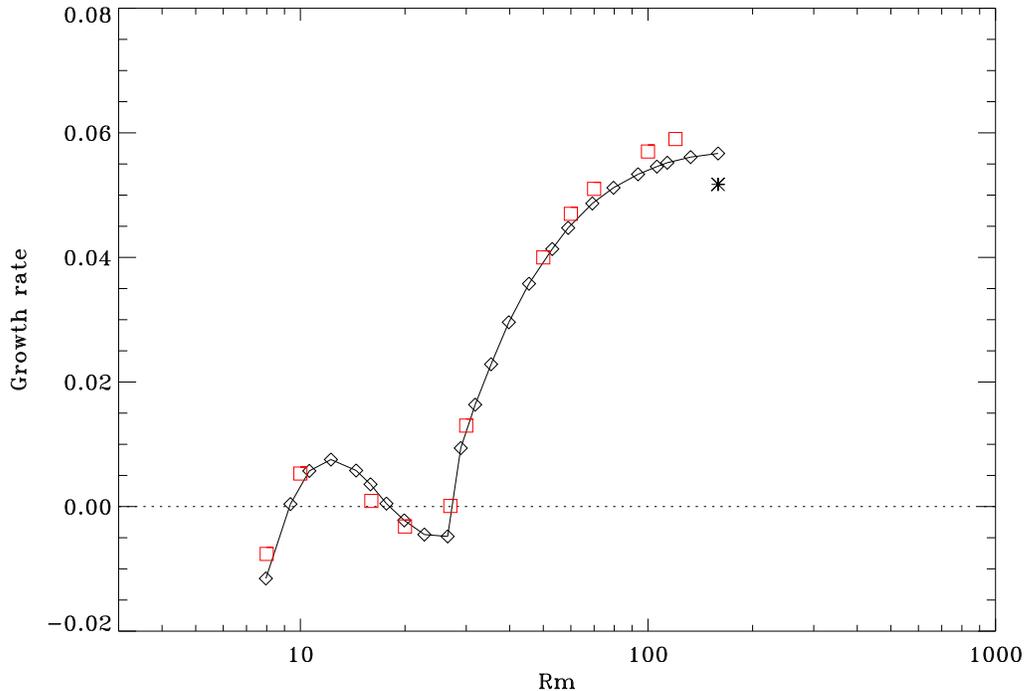}
\caption{Growth rate for the ABC dynamo as a  function of the magnetic
Reynolds number. This diagram agrees  remarkably well with the results
obtained  using   a  spectral  description by  \citeauthor{Galloway86}
\citeyear{Galloway86} (shown as boxes). The star is obtained with the AMR
implementation.}
\label{growthABC}
\end{figure}

\begin{figure}
\centerline{
\epsfxsize 9cm
\epsffile{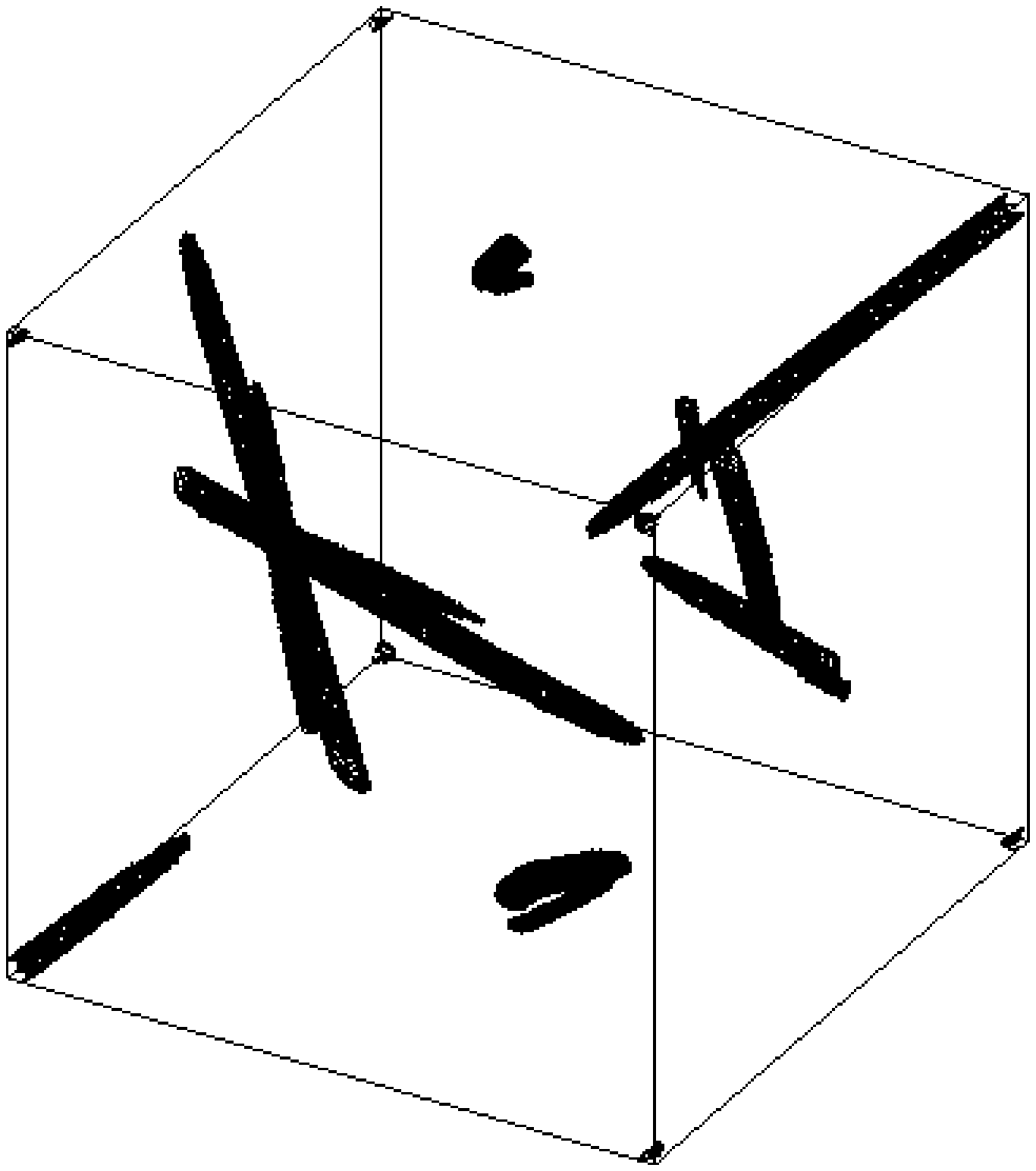}
\hskip -2cm
\epsfxsize 9cm
\epsffile{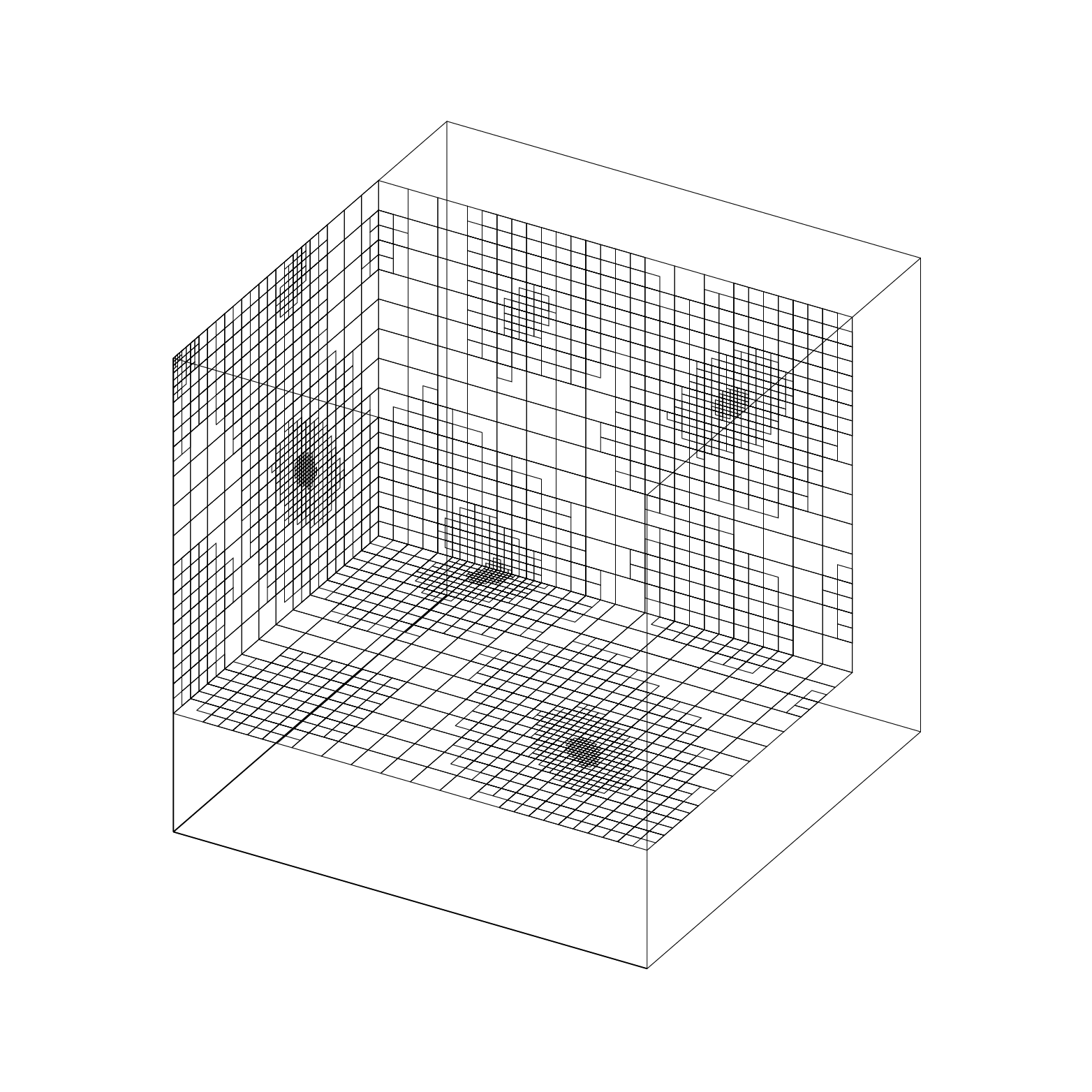}}
\caption{ABC dynamo investigated with  the AMR strategy  at $\Rm=159$.
On the left panel: surface of isovalue of  the magnetic energy density
$B^2/2=3\times 10^{19}$  at time $t=80$; on  the  right panel: the AMR
mesh geometry (for clarity, only ``octs'' boundaries are displayed here).}
\label{figabcamr}
\end{figure}

\begin{figure}
\vskip -5mm
\centerline{\epsfxsize \textwidth
\epsffile{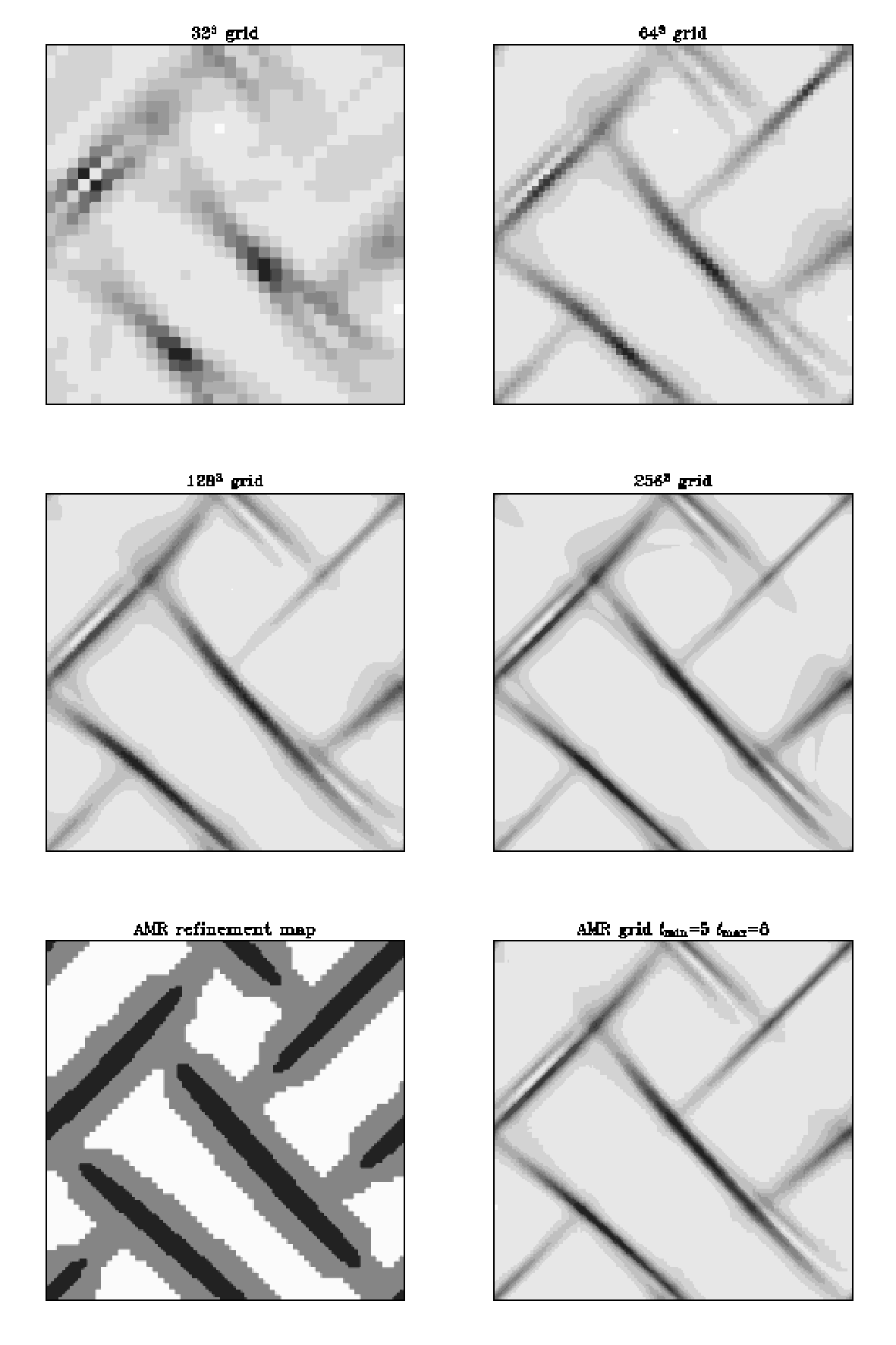}}
\vskip -5mm
\caption{The ABC  dynamo  is  investigated at $\Rm=159$   with various
resolutions.  The projected magnetic energy density is represented for
each run. The convergence  is demonstrated on  the Cartesian grid  and
the ability of the AMR grid to capture the solution is assessed.}
\label{figabcresol}
\end{figure}

This flow is  known  as a fast-dynamo:  at  large,  but finite,  $Rm$,
eigenmodes  in the form   of  cigar-shaped structures  develop  (e.g.
\citeauthor{gilbert},  \citeyear{gilbert}). They are very localized in
space (again $\ell  \sim  \Rm ^{-1/2}$), therefore constituting  ideal
candidates  for   a  investigation     using the   AMR    methodology.
Traditionally, these problems have been modeled using spectral methods
(e.g.  \citeauthor{Galloway86},  \citeyear{Galloway86}). The choice of
the velocity profile  in the form of Fourier  modes was largely guided
by the underlying  numerical method. More recently, \cite{Archontis03}
have  investigated this flow using  a staggered  grid and array valued
functions.

We want to emphasize here that because we are now investigating dynamo
action at large $\Rm$, the stability  properties of the Godunov scheme
will be   essential.   This will  be particularly  true   using an AMR
grid.  The  refinement strategy    will   ensure that   the   physical
resistivity dominates  on the finer grid which  is centered around the
cigar shaped magnetic structures (using  a threshold on $\mid\mid  \B
\mid\mid$).  Regions  relying on  a  coarser  grid,  however, will  be
dominated by the numerical  resistivity. The properties of the scheme,
both  in  terms  of stability and  of   low  numerical resistivity are
therefore essential ingredients to the success of the AMR methodology.

%

Dynamo action associated with this  flow is not  at all trivial. There
are at least two  regions of instability  in the parameter  space, one
for $8.9  \leq  \Rm \leq 17.5$ and  a  second for  $\Rm \geq  27$ (see
\citeauthor{Galloway86},     \citeyear{Galloway86}).       This second
instability has been followed up  to $\Rm$ of  a few thousand. We plan
to use our  methodology to investigate higher values  of $\Rm$  in the
near  future. This intricate  behavior of  the  growth rate with $\Rm$
suggests the use of high enough values of the magnetic Reynolds number
for convergence  study.   Otherwise, an increase of   the resisitivity
(decrease in $\Rm$)  could  yield an  increase  in the growth  rate by
sampling different regions of instability.

As in  the case of  the Ponomarenko dynamo,  we have calculated  the growth
rate as  a function of $\Rm$.   The corresponding graph,  using a Cartesian
grid     with     $(N_x,N_y,N_z)=(128,128,128)$     is     presented     on
figure~\ref{growthABC}~.  This  diagram is in excellent  agreement with the
spectral  results of \citeauthor{Galloway86},  \citeyear{Galloway86}, shown
in the same figure as squares.

We now  investigate this dynamo  using the AMR  scheme.  We want  to stress
that using AMR without care for such problems is not free of risk, the grid
being  affected by  the solution  and vice  versa.  Although  for  both the
advection and Ponomarenko tests, the  solution has been well captured using
straightforward refinement  criteria, the situation is more  subtle for the
ABC flow, for which the field  generation is not localized. If the strategy
is not  adequate, some  regions of the  flow might  not be refined  as they
should be, and thus be subject  to a large amount of numerical diffusivity.
The choice  of the optimal refinement  strategy for the ABC  flow is beyond
the scope of  the present study.  It could for example  be based on various
flow properties, such as Liapunov  exponents, stagnation points, etc, or on
various field properties, such as gradients, truncation errors, etc.

As a first step, we have used here a criterion based on the magnetic energy
density which  allows the grid to  be easily densified  near the cigar-like
structures: when the  local magnetic energy density on level  5, 6, 7... is
respectively greater than 4, 16,  64...  times the mean energy density, new
refinements are triggered. This strategy is best applied at large $\Rm$ for
which the magnetic structures are well localized. We focus here on $\Rm=159
\,\,(=1000/2 \pi)$.

The  AMR simulation yields  a growth  rate of  $0.052$ after  $77$~hours of
wall--time computing using $8$~processors.  It is evolved until $t=80$.  At
that time,  the grid  is composed of  $455659$~cells. The structure  of the
eigenmode   and   the   topology   of   the   grid   are   illustrated   in
figure~\ref{figabcamr}.  For comparison, the Cartesian grid simulation with
$256^3$ cells yields  a growth rate of $0.055$  but requires $138$~hours to
evolve the  solution only up to  $t=46$ and using  $64$~processors! The AMR
simulation has therefore allowed a gain  in memory of a factor of $37$, and
a  speed-up  of  $25$  in  time.  All  our  computations  are  compared  on
figure~\ref{figabcresol}.   The  first   four  panels  show  the  projected
magnetic   energy  obtained   varying   the  resolution   from  $32^3$   to
$256^3$. Computations performed with  $128^3$ and $256^3$ cells reveal very
little  differences  and  clearly  indicate convergence.   The  two  bottom
snapshots  illustrates the  structure of  the grid  in the  AMR simulations
({\it left panel})  and the projected magnetic energy  ({\it right panel}).
There is a good agreement between  the AMR  simulation and  the run  
performed  on the $256^3$ grid (about $10\%$). 


\section{Conclusions and perspectives}

We  have shown that the  Constrained Transport approach for preserving
the solenoidal character of the  magnetic field could be combined with
a Godunov  method, provided a two-dimensional   Riemann solver can be
used. We have  further shown how this  could be combined with  a MUSCL
high order scheme. We   considered  three schemes for  the  predictive
step, each with its own merits. For a uniform velocity field, these CT
schemes are strictly equivalent to well known finite volume schemes on
the staggered grid.  This important result provides additional support
to the advection properties of the CT framework.

We have  implemented this strategy  on a kinematic dynamo problem, for
which only  the induction  equation  needs   to be considered. We have
shown that the Godunov framework  allows an efficient AMR treatment of
fast dynamos,  by ensuring the  numerical stability  of the scheme  in
regions  solved with a coarse  grid (for which the effects of the 
physical diffusion are vanishing).

The approach introduced here  clearly needs to  be adapted to the full
set  of  MHD  equations, for   which  solving the Riemann problem is no longer
a trivial task. This important step  raises several  additional difficulties
and is the object of a forthcoming paper (\cite{Fromang06}).

\section*{Acknowledgments} 

We wish to thank St\'ephane Colombi for many useful discussions and
for help with the OpenMP implementation.
We are most grateful to Patrick Hennebelle for suggesting the use
of the C-MUSCL scheme.
Computations were performed on supercomputers at CCRT 
(CEA Bruy\`eres-le-Ch\^atel), DMPN (IPGP), and 
QMUL-HPCF (SRIF).
\bibliography{mhd}

\end{document}